\documentclass[fleqn,usenatbib]{mnras} 

\usepackage{microtype}
\usepackage[T1]{fontenc}

\DeclareRobustCommand{\VAN}[3]{#2}
\let\VANthebibliography\thebibliography
\def\thebibliography{\DeclareRobustCommand{\VAN}[3]{##3}\VANthebibliography}

\usepackage{graphicx}	% Including figure files
\usepackage{amsmath}	% Advanced maths commands
\usepackage{array}
\usepackage{booktabs}
\usepackage{float}
\usepackage{units}
\usepackage{newtxtext,newtxmath}
\usepackage{orcidlink}
\usepackage{soul}

\sethlcolor{yellow!80}  % Set highlight color
\title[\textsc{sponchpop}]{\textsc{sponchpop}: Population synthesis to investigate volatile sulfur as a fingerprint of gas giant formation histories}

\author[A. Sommerville-Thomas et al.]{
Anna Sommerville-Thomas$^{1}$\thanks{E-mail: annaleice.thomas.22@ucl.ac.uk}\orcidlink{0009-0005-1289-4835},
Mihkel Kama$^{1,2}$\orcidlink{0000-0003-0065-7267}, 
Oliver Shorttle$^{3}$\orcidlink{0000-0002-8713-1446},
and Jason Ran$^{1}$\orcidlink{0009-0006-6059-0841}
\\
$^{1}$ Department of Physics and Astronomy, University College London, UK \\
$^{2}$ Tartu Observatory, University of Tartu, Estonia \\
$^{3}$ Institute of Astronomy, University of Cambridge, UK
}

\date{Accepted XXX. Received YYY; in original form ZZZ}

\pubyear{2026}

\begin{document}
\label{firstpage}
\pagerange{\pageref{firstpage}--\pageref{lastpage}}
\maketitle

\begin{abstract}
Planet population synthesis is an integral tool for linking exoplanets to their formation environments. Most planet population synthesis studies have focused on the carbon-to-oxygen ratio (C/O) in gas or solids, yet more insight into planet formation may be afforded by considering a wider suite of elements. Sulfur is one such key element. It has been assumed to be entirely refractory in population synthesis models, restricting it to being a tracer of accreted rocky solids. However, sulfur also has a volatile reservoir dominant at the onset of star and planet formation, which is then converted into refractories. We investigate sulfur's wider potential as a formation history tracer by implementing a gas-grain chemical conversion, the first multi-phase treatment of S in a planet population synthesis model. We also present the planet formation module of \textsc{sponchpop} and its first predicted planet growth tracks and populations. We apply these to explore the diversity of the planetary sulfur budget. We show that planets can inherit a wide range of core and envelope sulfur content, depending on their formation environment and accretion history including late-stage infall, demonstrating sulfur's new potential as a diagnostic tool for planet formation. Our models predict that some rocky planets are born sulfur-poor, which may have significant implications for their geochemistry and habitability. Enhanced sulfur abundances in gas-giant atmospheres, such as in our solar system, may result not only from accretion of rocky planetesimals, but also from formation beyond the $\mathrm{H_2S}$ iceline.
\end{abstract}


%\texttt{TO DO:}
%\begin{itemize}
%\item Regenerate plots as pdfs for vectorising
%\item Figure 5 variants for: pebble drift on, pure viscous case, MHD wind case, 1e-3 alpha, varied fragmentation velocity, dtg ratio,
%\item Plots for same birth location but varied birth time
%\item Use mentat for full population  (50 final points on mass/semi major axis plot with sulfur mass fracs)
%\item finish type 2 migration prescription description
%\item Variance of planetesimal/pebble ratio
%\end{itemize}

\section{Introduction}

As of writing, just under 6000 exoplanets have been discovered, the diversity of which far exceeds that of the planets of our own Solar System. The properties of these exoplanetary bodies, their masses, semi-major axes, chemical compositions, and the systems they are part of all contribute to the understanding of their formation processes. While these formation processes may be universal, diversity arises from the differences in starting conditions, such as the birth location or time, host star properties, initial disk conditions, and encounters with other planets in the same system. Planet population synthesis modelling can bridge the gap between observations of protoplanetary disks and the fully formed planetary systems seen today \citep[e.g.,][]{mordasini2009, benz2014, chambers2018, turrini2021tracing}.

Studying the chemical composition of planet forming environments and the bulk chemical abundances of the the atmospheres of gas giants, enables us to potentially link protoplanetery disks to the planets they produce. The sulfur-, phosphorus-, oxygen-, nitrogen-, and carbon-bearing species along with hydrogen (SPONCH), when observed in the atmospheres of giant planets, can provide information on planets' atmospheric processes and provide valuable insights about possible formation histories \citep{oberg2011, turrini2021tracing, walsh2025}. A similar logic applies to the elemental budget of rocky planets, super-Earths, or giant planet cores.

Sulfur chemistry in protoplanetary disks represents a fundamental yet poorly understood component of planet formation processes. It is one of the most abundant elements in the universe (with an ISM value of S/H $\approx 1.32 \times 10^{-5}$ \citep{asplund2009}). Changes in the volatile or refractory budget of sulfur, such as through chemical kinetics, pebble drift, and other processes, may have implications for interpreting exoplanet compositions. Unlike carbon and oxygen, which have been extensively studied in the context of giant planet formation, sulfur exhibits greater chemical complexity. Sulfur has a wide range of possible redox states and forms species with a wide range of volatilities. These factors mean that sulfur couples to the dynamics of planet formation in more diverse ways than C and O, a behaviour that has not been captured by many previous population synthesis models. 

\begin{figure*}
\includegraphics[clip=,width=0.85\linewidth]{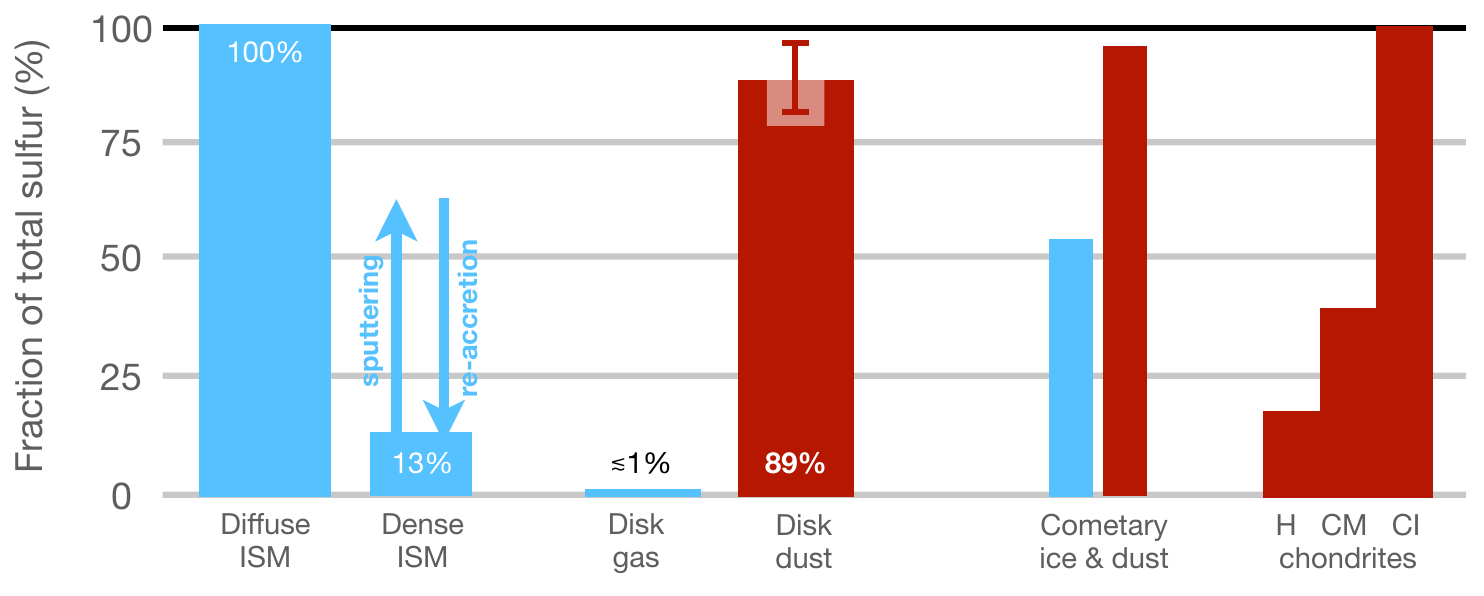}
\caption{Volatile (blue) and refractory (red) sulfur reservoirs as a percentage of total locally available sulfur, from the initial interstellar mass reservoir \citep{jenkins2009}, through protoplanetary disks \citep{fuente2010, dutrey2011, kama2019}, and as seen in solar system comets \citep{leroy2015, calmonte2016sulphur} and asteroids \citep{wasson1988compositions}. See \citet{kama2019} for full details.}
\label{fig:kama2019}
\end{figure*}

In the ISM, sulfur exists almost entirely as an atomic gas \citep{jenkins2009}, but in protoplanetary disks, it partitions between volatile molecules (e.g. $\mathrm{H_2S}$, CS, SO) and predominantly refractory condensates (such as FeS, MgS) \citep{semenov2018, kama2019}. In Figure\,\ref{fig:kama2019}, we show the evolution of sulfur from a nearly ideal volatile in the diffuse interstellar medium to its almost entirely refractory reservoirs in rocky planetesimals. In this paper, we employ a chemical kinetics model to track the processing of sulfur from a volatile gas into a refractory mineral as a function of time and local physical conditions, without invoking an equilibrium condensation assumption. We suggest this is consistent with evidence from protoplanetary disks \citep{fuente2010, dutrey2011, kama2019} and from solar system comets \citep{leroy2015, altwegg2016}, which reveals that in addition to a large refractory S reservoir, a significant volatile S reservoir may persist in particular in the outer disk where e.g. cometary ices originate.

Current population synthesis models typically assume 100\,$\%$ of sulfur resides in solid condensates \citep{turrini2021tracing, pacetti2022}, making atmospheric sulfur abundances direct tracers of accreted planetesimal/pebble (solid) material and planetary metallicity. However, this assumption is inconsistent with routinely detected sulfur bearing species (CS, SO, $\mathrm{H_2S}$) observed in protoplanetary disks by ALMA and Northern Extended Millimeter Array (NOEMA) \citep{phuong2018, booth2023sulphur, keyte2024}. Studies of Herbig\,Ae/Be systems reveal that as much as $\approx11\,$\% of all S atoms are in volatile form \citep{kama2019}, while the diffuse ISM results described above indicate that prior to molecular cloud formation almost all sulfur must be in an atomic gas \citep{jenkins2009}. While sulfur will ultimately exist as gas in hot planetary atmospheres regardless of its accretion phase, the bulk sulfur abundance of a planet's core or atmosphere may vary depending on the relative importance of sulfur inherited as gas, ice, or refractories during different stages of core or envelope accretion. This distinction is preserved in the total atmospheric abundance and can possibly reveal formation pathways.

%Recent observations of protoplanetary disks with ALMA have revealed significant depletion of gas-phase sulfur species in disk mid-planes \citep{legal2021}, suggesting efficient conversion between volatile and refractory forms that varies spatially and temporally. 

Hot Jupiters represent ideal laboratories for testing sulfur chemistry models due to their extreme atmospheric conditions and observational accessibility. Their close-in orbits result in atmospheric temperatures exceeding $1000\rm K$, where sulfur-bearing species achieve chemical equilibrium and become detectable through transmission and emission observations \citep{hobbs2021}. Critically, at these temperatures, sulfur species remain in the gas phase above the photosphere, making atmospheric abundances observed in transmission spectroscopy a reliable proxy for the complete atmospheric and envelope inventory of sulfur \citep{fu2024}. This absence of sulfur condensation at observable pressures distinguishes hot Jupiters from cooler gas giant planets where sulfur may condense below the photosphere, obscuring bulk composition. 

JWST's infrared capabilities have already begun revealing substantial sulfur in both Hot Jupiter and Neptune atmospheres that cannot be explained by refractory-only models. The detection of sulfur dioxide in WASP-39b's atmosphere by JWST represents a breakthrough in exoplanet atmospheric chemistry \citep{alderson2023, tsai2024, powell2024}. JWST has also detected hydrogen sulfide ($\mathrm{H_2S}$) in HD189733b's atmosphere, providing the first direct measurement of reduced sulfur species in an exoplanet \citep{fu2024}, and sulfur dioxide in warm Neptune WASP-107b \citep{dyrek2024}. These discoveries demonstrate that several reservoirs of sulfur in exoplanet atmospheres are detectable, and may provide new constraints on models of planet formation and atmospheric chemistry. The Ariel mission may eventually substantially expand the sample of exoplanets thus chemically characterised \citep{tinetti2022}.

These discoveries extend to our Solar System, where our understanding of Jupiter's atmospheric sulfur remains incomplete despite decades of study. Jupiter's atmosphere contains water, methane, hydrogen sulfide, ammonia and phosphine, with deep tropospheric measurements implying enrichment in carbon, nitrogen, and sulfur \citep{bolton2017}. The contrast between Jupiter's hidden sulfur chemistry and clear exoplanet detections highlights the fact that Jupiter's thermodynamics presents a distinct barrier to constraining the abundance of condensible species in it's atmosphere.

The key insight is that hot planets are necessary for using sulfur as a formation tracer, but the question is whether these hot giant planets formed cold and migrated inward or formed hot in-situ. Gas giants formed cold may have accreted beyond the $\mathrm{H_2S}$ snow line or migrated outwards, where volatile sulfur species would have been available for accretion, potentially leading to different sulfur contents compared to their warmer counterparts. The effect of formation in hot or cold regions of the disk for gas giant sulfur abundances can therefore be used as a possible indicator of formation location and migration scenarios, yet systematic studies incorporating realistic sulfur chemistry remain absent from literature. 

%Cold gas giants, while potentially enriched in sulfur through accretion of sulfur rich ices in the outer disk, present fundamental observational challenges as sulfur condenses out below the photosphere, making their bulk compositions difficult to constrain. This is consistent with our Solar System, where even Jupiter's deep atmospheric composition remains elusive.

In addition to tracing formation location, the atmospheric sulfur content of a hot giant planet may reflect complex interactions between atmospheric processing, chemical evolution, and late accretion to a planet.  These processes could also vary systematically between planet formation locations. Population synthesis approaches offer a natural framework for bridging this gap by incorporating sulfur chemistry into planetary formation and migration models.

Late-stage planetesimal infall has been thought to be a predominant mechanism for metal enrichment of planet atmospheres \citep{venturini2016}. Planets that do not have planetesimals ablate in their atmospheres in the later stages of their formation should instead see envelope metallicity that is comparably metal-poor, as they have only accreted gas which is, in essence, metal-poor.

In this paper, we explore the diagnostic power for tracing planetary formation history gained by accounting for the volatile-refractory chemical conversion of sulfur. Using the population synthesis code \textsc{sponchpop} and its planet formation module, we explore the interplay between sulfur chemistry and planet formation processes to provide these fingerprints of formation in giant planet atmospheres.  

\section{sponchpop}
\label{sec:sponch}

%popsynth goals and sponchpops specific goals, eg fully deterministic and computationally inexpensive
%what do we require the planet formation module for? how will it achieve this?
%intro to sponchpop and its science goals- linking stable planetary systems to their formation conditions, origins of planetary chemical diversity B) 
%initial conditions e.g. single run, runs for 3Myrs/stops at planet inner edge
%maybe separate section for initial conditions? to go through parameter table in a little more detail
%description of coupling between different physical processes, eg boundary conditions for core to gas accretion, boundary between type one and type two migration, and coupling of disk model to planet formation processes, eg solid surface density decrease for pebbles and planetesimals

The population synthesis model presented in this work, \textsc{sponchpop}, is based on multiple contributions to the field that aim to utilise different theories of core formation, accretion and planetary migration. It is a modular, fully analytical, Python-based code. This description of the model primarily focuses on the formation of the planet, and its interaction with its evolving host disk, computing and tracking the birth of a single planet for a $3\rm Myr$ duration. In this study, we aim to introduce \textsc{sponchpop} and use it to probe the importance of the remaining volatile sulfur budget to gas-giant formation.

\begin{figure*}
    \centering
    \includegraphics[width=1\linewidth]{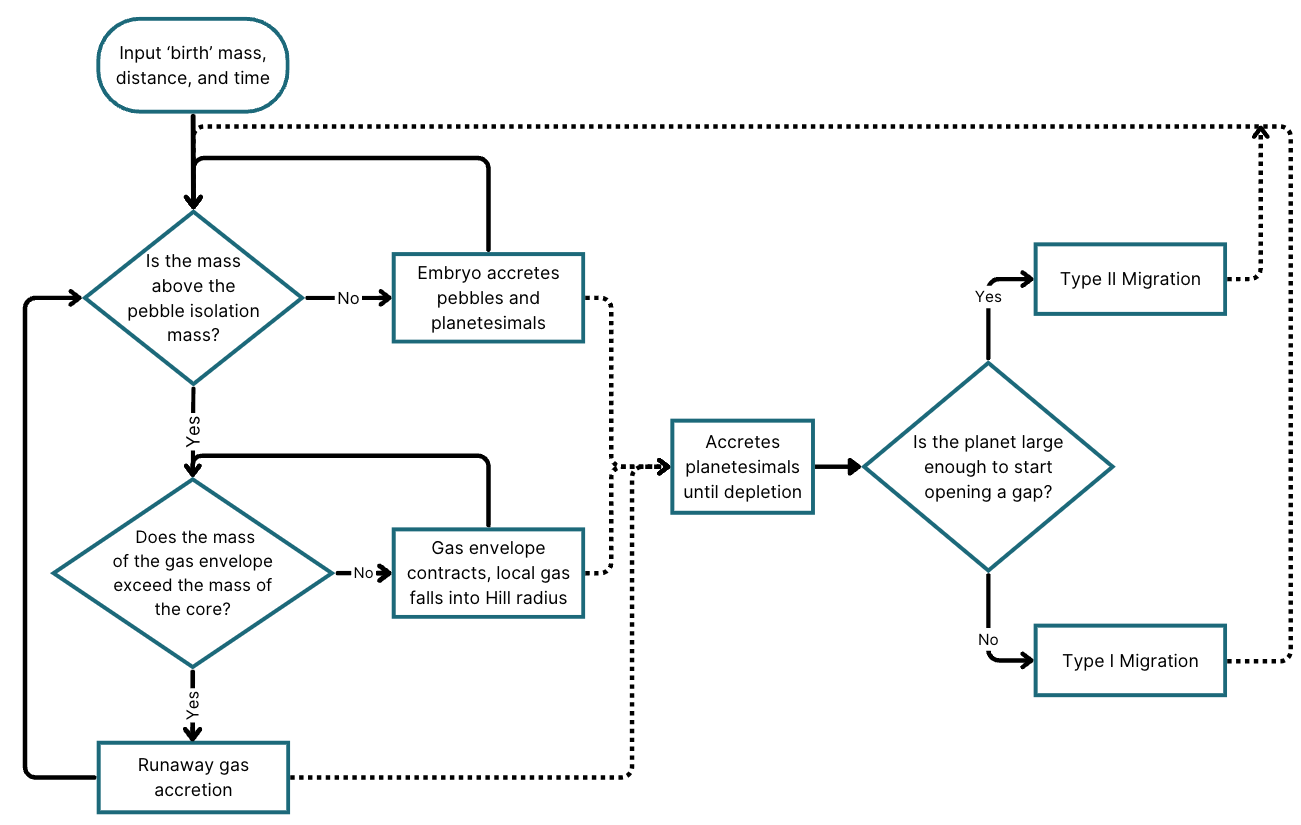}
    \caption{Flowchart showing the process undertaken by the fiducial case of the planet formation code.}
    \label{fig:flowchart}
\end{figure*}

During the formation phase of the planet, the core grows following the evolution of the disk. A static disk of planetesimals, as well as a surface density of pebbles contribute to the solid growth phase of the planet, the initial ratios of which are free parameters in this model. A gaseous envelope is bound to the core if the core is significantly massive (i.e., the isolation mass being reached), the contraction of which follows a Kelvin-Helmholtz timescale until the crossover mass is reached \citep{piso2014}, and runaway gas accretion can begin. These processes are described in depth in Section \hyperref[sec:pfm]{2.2}, and the progression of processes are shown in in Figure \hyperref[fig:flowchart]{2}.

\begin{table*}
\fontsize{8pt}{8pt}\selectfont
\centering
\begin{tabular}{ p{5cm} p{1cm} p{2.65cm} p{2.64cm}}

\multicolumn{1}{l}{\textbf{Parameter}} & \multicolumn{1}{l}{\textbf{ }} & \multicolumn{1}{l}{\textbf{Value/Range}} & \multicolumn{1}{l}{\textbf{Fiducial Value}}\\ \midrule
Disk alpha parameter & $\alpha$ & $1\times10^{-5} - 3\times10^{-3}$  & $1\times10^{-3}$\\
Dust to gas ratio & $f_{\rm dtg}$ &  $0.005-0.02$ & $0.01$\\
Initial outer disk radius & $s_0$ & $30-100AU$ & $35AU$ \\
Initial Embryo Mass & $M_{\rm p,0}$ & - & $10^{-5}M_{\oplus}$\\
`Birth' time & $t_{0}$ & $0-1Myr$ & $0.1Myr$\\
`Birth' Semi-Major Axis & $a_{0}$ & $0.05 - r_{\rm out, 0}$ & - \\
Planetesimal-to-solid ratio & $f_{\rm ptd}$ & - & $0.1$\\
Fragmentation velocity & $v_{\rm frag}$ & 2.5 $ms^{-1}$ & -\\
Stellar mass & $M_{\star}$ & $1 M_{\odot}$ & $1 M_{\odot}$ \\
Initial disk mass & $M_{d,0}$& $0.01 - 0.1 M_{\odot}$ & $1 \times 10^{-1} M_{\odot}$\\
Disk wind strength* & $f_{\rm w}$& $0 - 1$ & $0.8$\\
Mass loss scaling parameter* & $K$ & $0 - 1$ & $1$\\
\end{tabular}
\caption{Model parameters. Fiducial values are in the last column. Parameters with an * denote those only apply to the MHD wind disk model.}
\label{table:1}
\end{table*}

\begin{table}
\centering\begin{tabular}{lllllllll}
\hline
\multicolumn{1}{l}{\textbf{Species}} & \multicolumn{1}{l}{\textbf{Mass (amu)}} & \multicolumn{1}{l}{\textbf{$E_{des}$ (K)}} & \multicolumn{1}{l}{\textbf{$T_{sub}$(K)}} & \multicolumn{1}{l}{\textbf{S/H}}\\ \midrule
 %CO & 28 & 1150 & & $1.5\times10^{-4}$ & $0.0$ & $1.5\times10^{-4}$ & $0.0$ \\ 
 %CO2 & 44 & 2650 & &$3\times10^{-5}$ & $0.0$ & $6.0\times10^{-5}$ & $0.0$  \\
 %H2O & 18 & 5700 & &$0.0$ & $0.0$ & $9\times10^{-5}$ & $0.0$ \\
 %SiO4 & 136 & 64000 & &$0.0$ & $0.0$ & $1.47\times10^{-4}$ & $0.0$ \\
$\mathrm{H_2S}$ & 34 & 2700 & 54 & $8.84\times10^{-7}$ \\
 FeS & 88 & 33000 & 655& $1.17\times10^{-5}$ \\
 Fe & 56 & N/A & N/A & $0.0$ \\
 %$N_{salts}$ & 63 & 10000 & 200 & $0.0$ & h_2$0.0$ & $0.0$ & $0.0$ \\
\end{tabular}
\caption{The sublimation conditions and sulfur content of the chemical reservoirs considered in this work.}
\label{table:2}
\end{table}

\subsection{Disk models \& initial conditions}
\label{sec:diskmodel}

The disk module in \textsc{sponchpop} allows for a variety of 1D, or 1+1D protoplanetary disk models to be implemented. In this work, we make use of 1) viscous and irradiated disks from \citet{chambers2009};  and 2) a viscous, irradiated disk under the influence of an magnetohydrodynamical (MHD) disk wind from \citet{chambers2018}. For the case of the MHD wind disk model, we use fiducial parameters consistent with a fast wind and negligible mass loss due to the winds.
In both the MHD wind and viscous irradiated disk models, the boundary between the inner disk (heated through viscous motion) and the flared outer disk (heated through stellar irradiation) are divided by a transition radius that moves inwards over time as the disk mass accretion rate declines.
All host stars considered in this study are have solar mass, radius and effective temperature. The initial elemental abundances for $\rm Fe$ and $\rm S$ are taken from \cite{asplund2009}.

The input parameters for the disk models are detailed in Table \ref{table:1}. We use a fragmentation velocity corresponding to relatively fragile grains across the disk, $v_{\rm frag}=2.5 \rm ms^{-1}$, in agreement with models and observations \citep{jiang2024, tong2025}. The upper limit for the alpha viscosity parameter is taken from \cite{rosotti2023}, as disk turbulence is most likely weaker than previously assumed in the field, and other processes may be responsible for observed mass loss, such as MHD winds. Initial disk mass is limited to $10\%$ of a Solar mass. Initial planetesimal fraction for the disk is fixed at $10\%$ of the disk mass, motivated by masses of planetesimals formed after protoplanetary disk build-up \citep{drazkowska2018}. The evolution of the midplane temperature and total surface density in each disk model using the fiducial parameters specified in Table \ref{table:1} are displayed in Figure \ref{fig:all_disk_profiles}.

\subsection{Planet formation model}
\label{sec:pfm}

The evolution of the bulk atmospheric composition of gas giants is intrinsically related to the formation history of the planet and the local conditions of its host disk. The planet formation model here is designed to be efficient, meaning assumptions have been made to enable the models to be computationally feasible.

\subsubsection{Core accretion}
\label{sec:coreacc}

In \textsc{sponchpop}, we consider the mass of the core of the protoplanet to grow via both pebble and planetesimal accretion. Core accretion via pebbles is sensitive to the conditions of the host disk. Grain growth, distribution and midplane settling can all be influenced heavily by the turbulence in the disk. In order to predict pebble size and limit pebble accretion before sufficient grain growth, \textit{pebble predictor} \citep{pebpre} is employed, a pebble flux and growth predictor motivated by \cite{twopoppy}'s \textit{two-pop-py}. A more in depth description of the models can be found in the the corresponding papers. This model was chosen to limit the time required to run \textsc{sponchpop}, as \textit{pebble predictor} does not utilise a fully time integrated protoplanetary disk model.

For \textsc{sponchpop}, \textit{pebble predictor}'s Stokes growth timescale is used to grow grains in the disk until the fragmentation or radial drift barriers are met. The model assumes that dust grains are micrometer sized monomers that aggregate in disk turbulence driven encounters, following the form of

\begin{equation}
\tau = f_{\rm dtg}  \frac{1}{\Omega_{\rm K}} (\frac{\alpha}{10^{-4}})^{-1/3} (\frac{r}{AU})^{1/3},
\end{equation}
meaning that the Stokes number grows exponentially as

\begin{equation}
St =  St_{0} \exp(\frac{t}{\tau})
\end{equation}
where $St_{0}$ is the initial Stokes number, defined as corresponding to micrometer sized monomers. While the process of grain growth is included in \textsc{sponchpop}, it has been excluded from this science case to simplify the chemical kinetics.

When $St$ exceeds a minimum value of $1 \times 10 ^{-3}$, the grains are considered pebbles and are able to be accreted onto the protoplanet embryo \citep{ormel2017}. The Stokes number is considered to be limited by one of two processes; radial drift to the host star and turbulence fragmenting pebbles into smaller grains. 

A single embryo of $10^{-5}M_{\oplus}$ is injected into the disk. Following this, pebble and planetesimal accretion take place simultaneously until the pebble isolation mass is reached, where pebble accretion ceases. Planetesimal accretion continues until the surface density of planetesimals in the protoplanet's feeding zone has been exhausted. For the duration of the gas accretion regime, planetesimals accreted beyond the pebble isolation mass contribute both to the mass of the core \textit{and} the mass of the gas envelope via planetesimal ablation, dependent on a free parameter $p_{\rm ratio}$, given in Table \ref{table:1}.

The initial surface density of solids in the disk follows the gas surface density such that $\Sigma_{\rm solid} = f_{\rm dg}\Sigma_{\rm gas}$, where $f_{\rm dg}$ is the dust-to-gas ratio. Of these solids, a fraction of the total mass is allocated to grains and pebbles, and a fraction is allocated to planetesimals. 

\textbf{Planetesimal accretion: }A planetesimal `disk' is initialised at $t_{0}$ of the disk, so that the surface density of larger bodies remain stationary between the boundary of the inner edge and the initial outer edge of the disk. The initial fraction of the solids allocated to the planetesimals is a free parameter, the value of which for our baseline model is shown in Table \ref{table:1}.  This implementation comes with the assumption that planetesimals are constant across the disk and available at $t_{0}$. Planetesimal formation can occur through the streaming instability mechanism in portions of the disk where the dust-to-gas ratio is enhanced by the presence of ice lines \citep{schafer2017, liu2019}, or even dust trapping caused by planet formation. Despite this, the planetesimal sulfur density implemented here does not include these local enhancements.

For accretion of planetesimals onto the injected embryo, a modified prescription from \cite{safronov1969} is followed such that

\begin{equation}
(\frac{dM_{\rm p}}{dt}) = \Omega \Sigma_{\rm pmal} R^{2}_{\rm pmalcap}v_{\rm rel} F_{\rm G},
\end{equation}
where $F_{G}$ is the two body gravitational focusing factor, a term describing the enhancement of collision rates due to gravitational effects given by $F_{\rm G} = 1 + (v_{\rm esc}/v_{\rm rel})^{2}$, and $R_{\rm pmalcap}$ is the capture radius of planetesimals given by $R_{\rm pmalcap} = r_{\rm p} F^{1/2}_{\rm G}$.

Unlike for the population of pebbles, which is replenished in the feeding zone of the protoplanet by the inwards drift of material, the surface density of planetesimals decreases in line with the amount of mass accreted by the planet. It is assumed that surface density is uniform across the feeding zone \citep{thommes2003}, and the decrease of solids is given by:
\begin{equation}
(\frac{d\Sigma_{\rm pmal}}{dt}) = - \frac{(3 M_{*})^{1/3}}{6 \pi a^{2}_{\rm p} B_{\rm L} M^{1/3}_{\rm p}} \frac{dM_{\rm p}}{dt},
\end{equation}

where $B_{\rm L}$ is the width of the feeding zone of planetesimals in Hill radii; and $r_{\rm hill} \approx a^{3} \sqrt{\frac{M_{\rm p}}{3M_{*}}}$. Planetesimal accretion continues until the local population of planetesimals is depleted by accretion onto the planet, or the simulation ends. 

The starting abundance of refractory sulfur across the planetesimals is initialised such that within the radius FeS sublimates, the planetesimals have a sulfur mass fraction of 0. Between the FeS sublimation radius and the location in the disk where temperatures are too cool for efficient FeS formation by chemical kinetics, planetesimals have a S mass fraction of $S/X$, where X accounts for the 18 most abundant refractory elements and roughly 20\% of both carbon and oxygen \citep{lodders2003}. From this location until the $\mathrm{H_2S}$ snowline, planetesimals once again hold no sulfur, and beyond the snowline they again have a sulfur mass fraction of $S/X$.

The free parameter dictating the mass fraction of accreted planetesimals contributing to the envelope mass after pebble isolation mass has been reached is varied across the three cases presented here: 

\begin{itemize}
\item[a)] $p_{\rm ratio}= 0.5$, our fiducial case, where planetesimal accretion during the runaway gas accretion phase leads to $50\%$ of accreted planetesimal mass (and the sulfur they carry) contributing to the core and envelope respectively
\item[b)] $p_{\rm ratio} = 1$, where it is assumed that planetesimals are entirely ablated in the planet's atmosphere when accreted during runaway gas accretion. In these simulations, all sulfur enrichment in the envelope is due to this planetesimal infall, and no volatile sulfur is accreted during the rapid gas accretion phase, to simulate disks where gas-phase sulfur has been entirely depleted
\item[c)] $p_{\rm ratio} = 0$, where accreted planetesimals only contribute to the mass and sulfur fraction of the core. In this case, sulfur enrichment of the envelope only occurs through accretion of volatile $\mathrm{H_2S}$ during envelope contraction and runaway gas accretion.
\end{itemize}

We also include a second model case of $p_{\rm ratio}=1$, where the gas-phase sulfur is included.

\textbf{Pebble accretion: }In the case of pebbles, the pebble surface density instead evolves as grains either drift inward towards the host star, or are locally accreted onto the planet and therefore removed from the disk. Following \cite{ormel2010}, the rate at which a planetary embryo accretes pebbles in both a 2D and 3D regime is,
\par

\begin{center}
\begin{equation} \label{eq:5}
\frac{dM_{\rm p}}{dt} =
\begin{cases} 
\frac{\pi R^{2}_{\rm cap}v_{\rm rel}\Sigma_{\rm peb}}{2 H_{\rm peb}}, & R_{\rm cap} < H_{\rm peb} \\[10pt]
2 R_{\rm cap} v_{\rm rel} \Sigma_{\rm peb}, & R_{\rm cap} > H_{\rm peb}
\end{cases}
\end{equation}
\end{center}
where $R_{\rm cap}$ is the capture radius of pebbles, $v_{\rm rel}$ is the relative velocity between a pebble and a planet, and $H_{\rm peb}$ is the scale height of the pebbles in the disk, dependant on the local Stokes number, turbulence and gas scale height. 

Pebble accretion continues until the planet reaches the pebble isolation mass, the mass at which the local gas disk is disturbed enough to create a pressure bump outside of the planet's location, effectively stopping the radial drift of pebbles $St << 1$. The pebble isolation mass is described using \cite{lambrechts2014}, where

\begin{equation}
M_{\rm iso} \approx 20 (\frac{H/R}{0.05})^{3/4} M_{\rm \oplus}.
\end{equation}

The disk model is structured as a logarithmic radial grid from the inner edge and ending at 250 AU. The radial drift of pebbles through each grid cell, $v_{\rm r}$ is given by \cite{w1997} as,

\begin{equation}
v_{\rm r} =  \frac{2 \eta v_{\rm kep} St}{1 + St ^{2}}
\label{eq:drift}
\end{equation}
where $\eta$ is the fractional difference between the sound speed $c_{\rm s}$ and the Keplarian velocity of the pebbles $\eta = (\frac{c_{\rm s}}{v_{\rm kep}})^{2}$. This equation of radial drift is used to describe how pebbles drift inwards towards the host star depending on the Stokes number. \textsc{sponchpop} uses a single population of pebbles across the disk, initialised as a fraction of the solids by a free parameter.

At $t_{0}$ of disk evolution, no FeS is present in the disk. A fraction of the small grains and gas are allocated to the initial Fe and $\mathrm{H_2S}$ respectively, the mass fraction of which calculated from \cite{asplund2009} as a fraction of the twenty most abundant elements in the present-day solar photosphere.

\subsubsection{Gas accretion}
\label{sec:gasacc}

Up until the pebble isolation mass is reached, we assume that the infall of pebbles through the protoplanet's preliminary atmosphere heats the gas enough to prevent collapse by the energy supplied by accretion. Though there is still planetesimal accretion taking place, it is at a reduced rate by the time $M_{\rm iso}$ is reached, as the local planetesimal population is not replenished and therefore the luminosity incident on the protoplanet during the gas accretion phase is not considered.

Accretion of a gas envelope (still attached to the gas disk) is assumed to have begun during the solid accretion phase of formation. By the time the pebble isolation mass is hit, 10\% of the total planet mass is allocated to a preliminary gas envelope, as we assume that the core accretion phase included the accretion of some volatiles. At this point, this envelope can begin to undergo contraction and cooling on a Kelvin-Helmholtz timescale. 

Envelope contraction follows the analytical prescription from \cite{piso2014}, describing the contraction of the gas envelope as a function of the protoplanet's core, using the formalism of \cite{bitsch2019}.

\begin{equation}
\begin{aligned}
\frac{dM_{\rm gas}}{dt} &= 0.000175 f^{-2} \left(\frac{\kappa_{\rm env}}{1 \,\text{cm}^{2}/\text{g}}\right)^{-1} \left(\frac{\rho_{\rm core}}{5.5 \,\text{g}/\text{cm}^{3}}\right)^{-1/6} \\ 
&\quad \times \left(\frac{M_{\rm core}}{M_{\rm env}}\right)^{11/3} \left(\frac{M_{\rm env}}{M_{\rm \oplus}}\right)^{-1} \left(\frac{T}{81 \,\text{K}}\right)^{-0.5} \frac{M_{\oplus}}{\text{Myr}}
\end{aligned}
\end{equation}
where $f= 0.2$ is a factor to describe the accretion rate to match those of numerical and analytical simulations \citep{piso2014}. $\kappa_{\rm env}$ denotes the opacity of the planet's envelope, given as $0.05\,\mathrm{cm^{2}/g}$, and $\rho_{\rm core}$ is the density of the solid core, taken as $5.51\,\mathrm{g/cm^{3}}$ (the average density of the Earth). $M_{\rm core}$ denotes the remaining $90\%$ of the mass at the time of $M_{\rm iso}$ is reached. This continues until critical mass ($M_{\rm core} = M_{\rm env}$) is reached.

After the mass of the gas envelope exceeds the mass of the core, runaway gas accretion can begin, which is truncated in this model by the dissipation of the gas disk in the later stages of evolution, or the ending of the simulation at $3\,\text{Myr}$. The mass of the preliminary atmosphere is a strong function of the solid accretion rate and the opacity of the gas. When the mass of the envelope has reached or exceeded the mass of the core, the gas accretion rate onto the planet follows

\begin{equation}
\begin{aligned}
\dot{M}_{\rm gas, low} &= 0.83 \,\Omega_{\rm K} \Sigma_{\rm g} H^{2} \left(\frac{r_{\rm H}}{H}\right)^{9/2} 
    && \quad \text{if } \frac{r_{\rm H}}{H} < 0.3, \\[8pt]
\dot{M}_{\rm gas, high} &= 0.14 \,\Omega_{\rm K} \Sigma_{\rm g} H^{2} 
    && \quad \text{if } \frac{r_{\rm H}}{H} > 0.3
\end{aligned}
\end{equation}
where the high and low mass branches of runaway gas accretion are a result from 3D hydro-dynamical simulations from \cite{machida2010}, which found different rates of runaway gas accretion from planets above a certain Hill radius to gas scale height ratio. In the case of both runaway gas accretion and envelope contraction, removal of gas from the protoplanetary disk is not considered. The surface density of the gas local to the planet only decreases with the evolution of the protoplanetary disk.

\subsubsection{Migration}
\label{sec:mig}

In the early stages of the planet formation process when the protoplanet is not yet, or never will, reach a mass significant enough to perturb the gas disk, it undergoes Type\,I migration, wherein the inwards drift of a disk-embedded planet is determined by the interaction between the disk and the planet. Inward migration in this case determined by an imbalanced torque from Lindblad resonances within the planet's orbit. The migration prescription utilised is from \cite{tanaka2002}. This formulation is used as it assumes the protoplanetary disk is locally vertically isothermal, which is the case for our disk models. Using the total net torque (Lindblad and co-rotation) for a 2D disk,

\begin{equation}
    \Gamma_{\rm total, 2D} = (1.16 + 2.828 \alpha_{\Sigma})(\frac{M_{\rm p}}{M_{\star}} \frac{a_{\rm p} \Omega_{\rm K}}{c_{\rm s}})^{2} \sigma_{\rm p}a^{4}_{\rm p} \Omega^{2}_{\rm K}
\end{equation}
where $\Omega_{\rm K}$ is the Keplarian velocity of the planet, and $\alpha_{\Sigma}$ is the gas surface density gradient local to the planet. In the case of the disk models used here, the surface density gradient is always negative, therefore the net torque is positive, resulting in migration that always acts inwards. The migration rate follows

\begin{equation}
    \dot{a}_{\rm p,1} = - 2a_{\rm p} \frac{\Gamma_{total}}{L_{\rm p}}
\end{equation}
where the angular momentum of the planet is $L_{\rm p} = M_{\rm p} (GM_{\star} a_{\rm p})^{1/2}$. This migration rate is applicable to planets that have not yet accreted enough to perturb the gas pressure gradient and start to carve a gap in the disk. This migration rate changes from Type\,I to Type\,II when the Hill radius of the planet exceeds the scale height $H_{\rm gas}$ of the disk \citep{alibert2005}. 

\begin{figure} 
\includegraphics[clip=,width=0.95\linewidth]{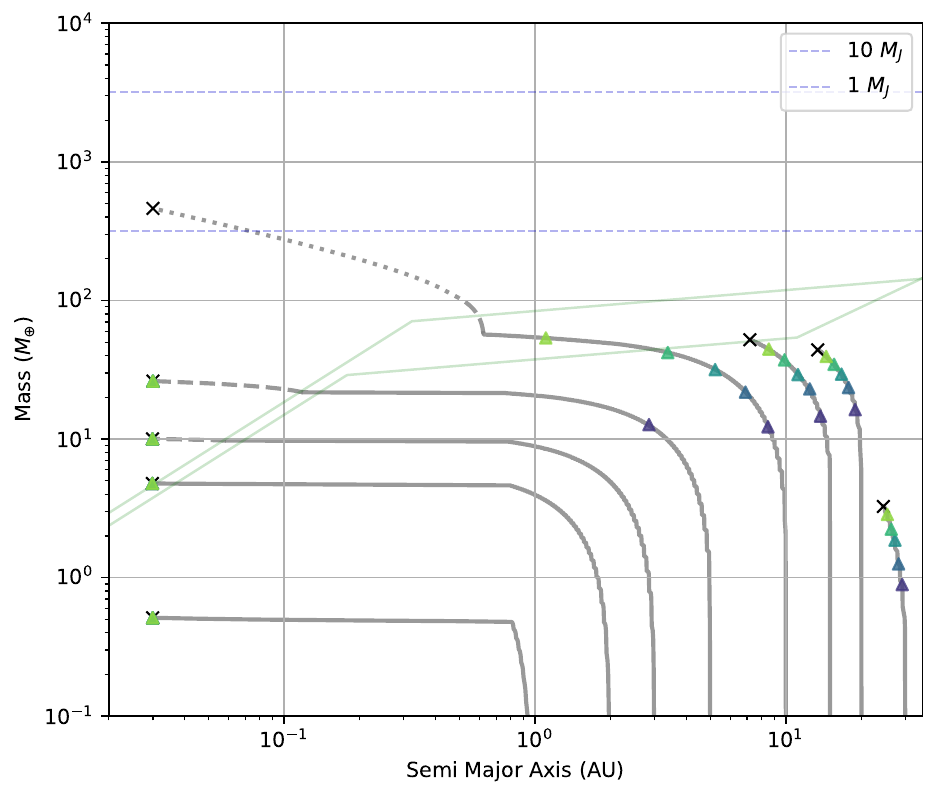}
\caption{Planet formation tracks for the fiducial parameter set detailed in Table \ref{table:1}, using the viscous-irradiated disk model. The solid, dashed and dotted parts of each track represent the solid accretion, envelope contraction and runaway gas accretion, respectively. The coloured triangles are markers for every $0.5\,\text{Myr}$ from $t_0$ of planet birth time, ending after $3\,\text{Myr}$. The green lines denote the minimum and maximum pebble isolation masses for this specific disk model.}
\label{fig:evotracks_nochem}
\end{figure}

For planets that meet that condition, Type\,II migration begins following \cite{scardoni2020}, expressed as

\begin{equation}
\dot{a}_{\rm p,2} = - u_{\rm r} \frac{B}{B+1}
\label{eq:mig2}
\end{equation}
where $u_{\rm r}$ is the gas radial velocity, which is dependant on the mass accretion rate of the protoplanetary disk. $B$ is a function of the planet mass

\begin{equation}
B = \frac{4\pi a^{2} \Sigma_{0}}{M_{\rm p}}
\label{eq:mig22}
\end{equation}
and $\Sigma_{0}$ is the unperturbed gas surface density at the planet's location. Type\,II migration is slower than Type\,I, where planets move inwards proportionally to the viscous speed of the disk in two different regimes. The earlier regime is dominated by the disk, and migration rates are closer to the gas radial velocity. The later regime of Type\,II migration, for very large planets, sees a significantly slower migration rate due to the planet's inertia. An example of growth tracks produced by the fiducial disk parameters (Table \ref{table:1}) is shown in Figure \ref{fig:evotracks_nochem}.

\begin{figure} 
\includegraphics[clip=,width=0.95\linewidth]{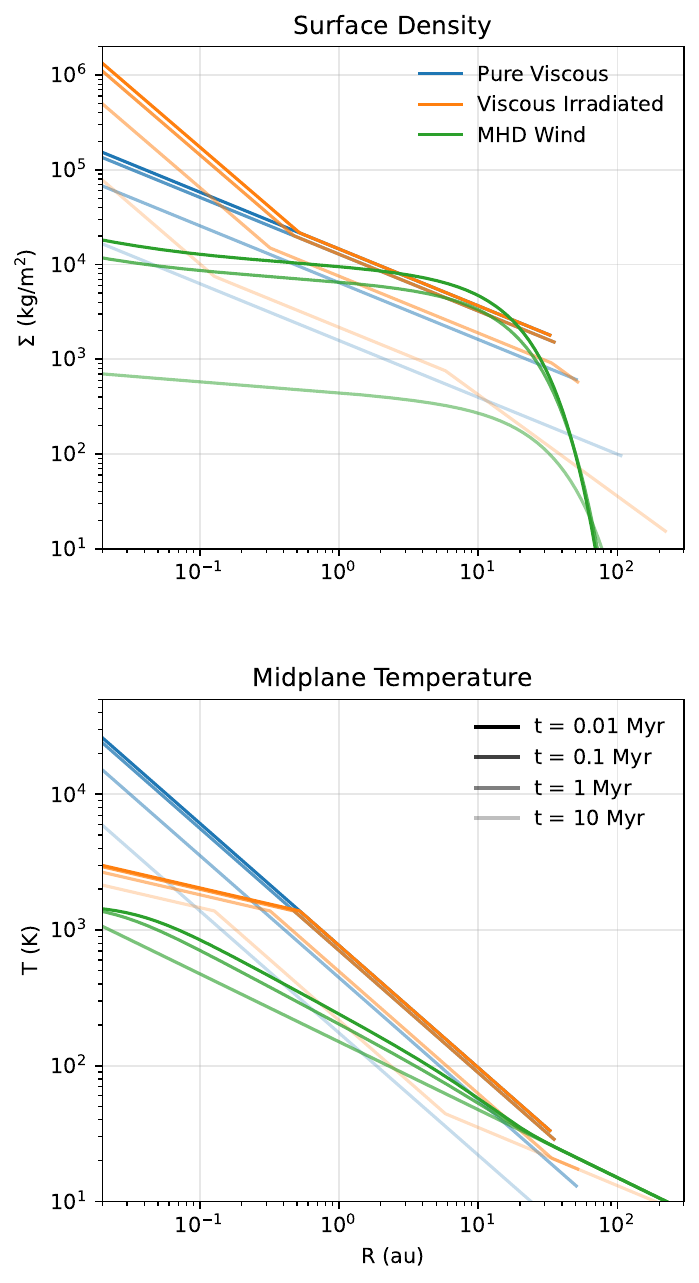}
\caption{Surface density distribution and temperature profiles of each considered disk model. Line colour denotes disk model, and opacity denotes time in disk evolution.}
\label{fig:all_disk_profiles}
\end{figure}

\begin{figure*} 
\includegraphics[clip=,width=1\linewidth]{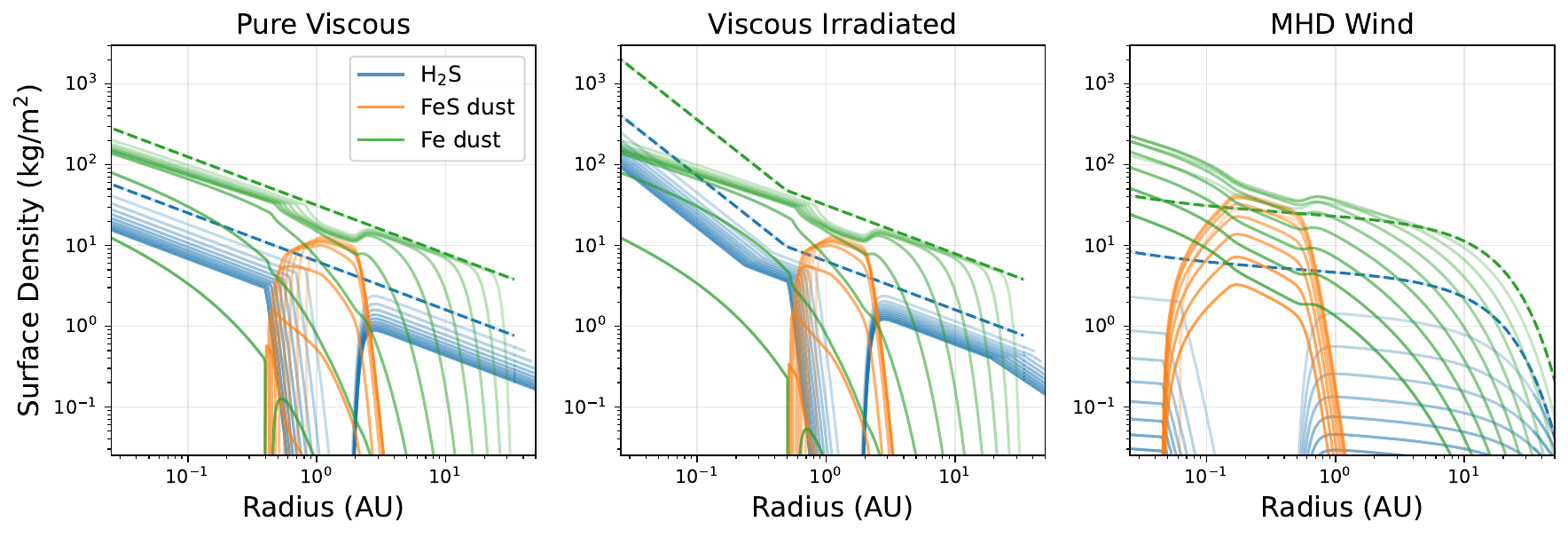}
\caption{Surface density distributions for the FeS, Fe and $\mathrm{H_2S}$ species in each considered disk model following $3\,\text{Myr}$ of disk evolution. The dashed lines show the surface densities at $t_0$ of the disk (only applicable for Fe and $\mathrm{H_2S}$, before any FeS is formed), and the solid lines denote the densities at $0.5\,\text{Myr}$. The more transparent the line, the earlier in the disk lifetime the distribution. These distributions are for disks sans planet formation, using the fiducial disk parameters outlined in Table \ref{table:1}}
\label{fig:species_evo}
\end{figure*}

\subsection{Chemistry}
\label{sec:chem}

We include a minimalistic chemical reaction network with the main aim of describing important reservoirs and phase transitions of sulfur.

Sulfur has been assumed to be entirely refractory in those planet formation-tracing and population synthesis models that have included it \citep[see e.g.,][]{oberg2019jupiter, turrini2021tracing}. Sulfur is, however, mostly or entirely present as an atomic gas in the diffuse interstellar medium \citep[][and references therein]{jenkins2009}, while in (extra-)Solar System solids rocky bodies generally carry all the sulfur \citep{wasson1988compositions, gansicke2012, Xuetal2013}. Comets contain a large amount of sulfur both in their dust and ices \citep{calmonte2016sulphur}.

One hypothesis for the conversion of volatile atomic sulfur into protoplanetary dust and rocks is the reaction of gas-phase sulfur-bearing molecules, such as H$_{2}$S or OCS, with iron-rich solids in a protoplanetary disk. We implement the linear kinetics of this following laboratory measurements by \citet{lauretta1996rate}:

\begin{equation}
\label{eq:h2s_to_fes}
{\rm H}_{2}{\rm S} + {\rm gFe} \rightarrow {\rm gFeS} + {\rm H}_{2},
\end{equation}

where ``g'' denotes a refractory (grain) phase. The reaction in Equation\,\ref{eq:h2s_to_fes} has the forward and reverse reaction coefficients, in units of g\,cm$^{-2}$\,h$^{-1}$\,atm$^{-1}$ of FeS produced or destroyed, given by

\begin{equation}
k_{\rm f} = (5.6\pm1.3)\,\exp{\frac{-27950\pm7280}{R\,T}}
\label{eq:H2StoFeS}
\end{equation}
and
\begin{equation}
k_{\rm r} = (10.3\pm1.0)\,\exp{\frac{-92610\pm350}{R\,T}},
\label{eq:FeStoH2S}
\end{equation}
where $R$ is the ideal gas constant and $T$ is the temperature of the system in Kelvin.

In our implementation, the H$_{2}$S gas is assumed to be in contact with pure iron with a total surface area representing the fraction of dust that is accounted for by Fe at solar abundances from \citet{asplund2009}. The grain size relevant to FeS formation is assumed to be a constant $a_{\rm Fe}=1\,$mm, typical for pebbles in the outer disk. As there is no complete dust population model yet in \textsc{sponchpop}, we have adopted this fixed $a_{\rm Fe}$ to balance grain growth --- which weights the solid mass towards increasingly larger particles --- with the dominant role smaller grains tend to play in the total area available for reactions. The gas and dust temperatures are assumed to be equal. Then, both the forward and reverse reaction rates are calculated, resulting in either the creation (Eq.\,\ref{eq:H2StoFeS}) or destruction  (Eq.\,\ref{eq:FeStoH2S}) of FeS. 

H$_{2}$S can exist either as a gas or ice in our model. For the thermal desorption of H$_{2}$S, we adopt a binding energy of $E_{\rm b,H_{2}S}=2700\,$K \citep{Wakelametal2017}, corresponding to a sublimation temperature of $T_{\rm sub,H_{2}S}\approx 50\,$K. This is substantially lower than the $\sim300$ to $400\,$K range where Eq.\,\ref{eq:H2StoFeS} converts mobile H$_{2}$S to FeS. Figure \ref{fig:species_evo} shows the species surface density evolution of all three for the disk models considered in this work when utilising our fiducial disk parameters (Table \ref{table:1}). 

\begin{figure*} 
\includegraphics[clip=,width=1\linewidth]{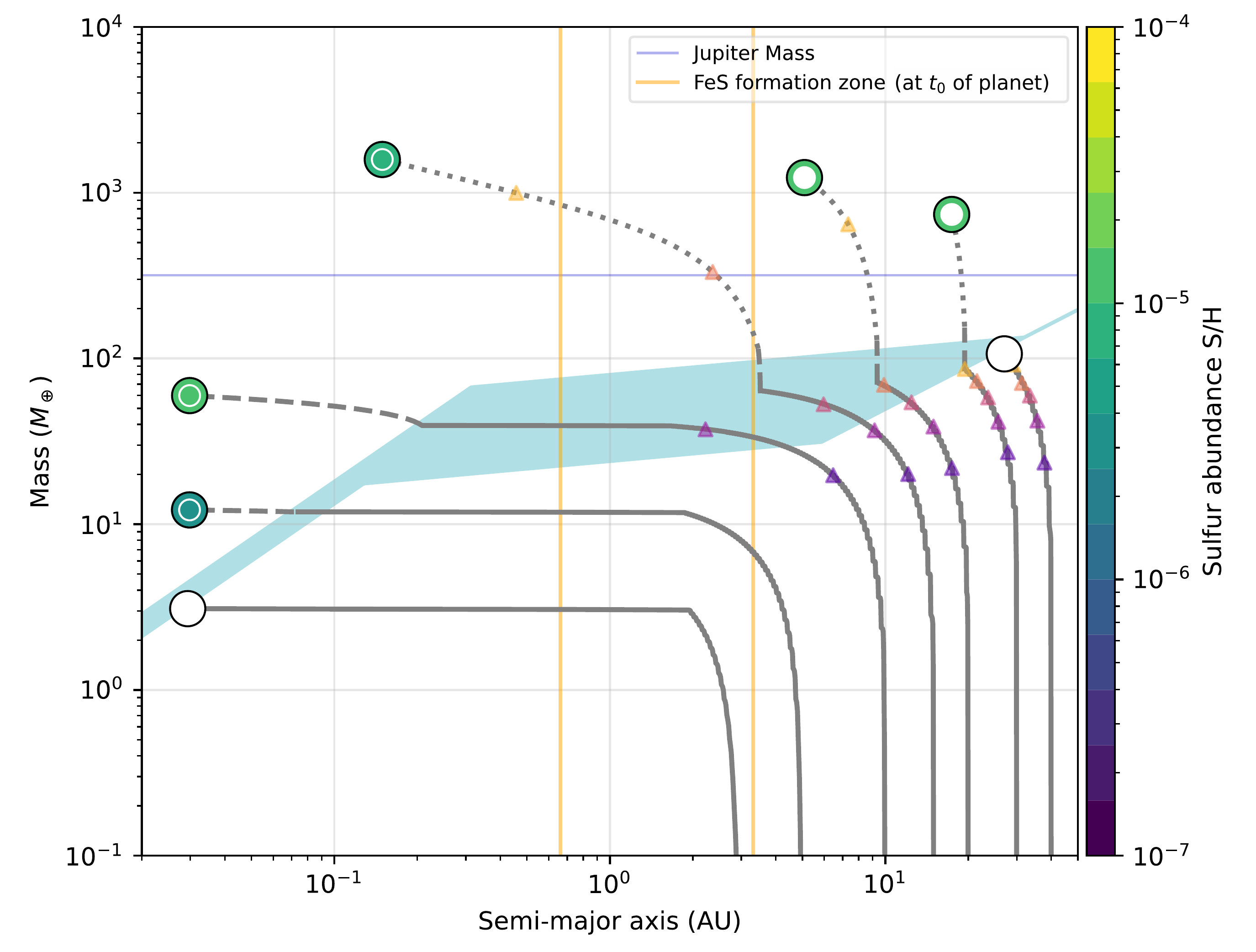}
\caption{Formation tracks of planets in the viscous irradiated disk with an alpha viscosity value of $\alpha=0.7\times10^{-3}$ and starting mass of $M_{\rm disk,0}= 0.075 M_{\odot}$. All other parameters are fiducial. The solid, dashed and dotted parts of each track represent the solid accretion, envelope contraction and runaway gas accretion, respectively. The marker colour denotes the final sulfur mass fraction in the envelope (outer ring) and core (solid centre) respectively, and planets with negligible sulfur content are marked in white (values below $10^{-10}$). The blue shaded region represents the minimum and maximum pebble isolation mass. Planet markers that are empty are planets that have no/negligible sulfur in both the core and envelope. This set of planets is generated with a $p_{\rm ratio}=0$, to demonstrate spatial dependence on sulfur abundance without supplementation from refractory sulfur reservoirs.}
\label{fig:chem_tracks}
\end{figure*}

\section{Results}

We utilise the planet population synthesis model, \textsc{sponchpop}, described in Section \ref{sec:sponch}, generating a total of 45,000 individual planets across three disk models and employing three planetesimal ablation ratios. We use a simple bi-directional gas-grain sulfur reaction to describe the main reservoirs of gas-phase and refractory sulfur in the form of pebbles and small grains throughout the protoplanetary disk, an example of which (without planet formation) is shown in Figure \ref{fig:species_evo}. The gap created in each disk profile produced by the efficient production of FeS is still significant after $10 \rm Myrs$ of evolution (Figure \ref{fig:h2s_sigma}). The host disk of each planet is varied between the disk models described in Section \ref{sec:diskmodel}, with initial disk parameters varied in the ranges specified in Table \ref{table:1}. 

We vary our ratio of planetesimal ablation in gas atmospheres, assuming a fiducial value of  $p_{\rm ratio}=0.5$, and two 'extreme' values of 0 and 1, representing negligible and total planetesimal ablation respectively. This was done to isolate sulfur abundances created by the presence of sulfur entirely in the gas phase and entirely in the refractory across the disk. In the case of our total planetesimal ablation case, $p_{\rm ratio}=1$, we assume that all sulfur is entirely in the solid phase and is carried by planetesimals, and that all gas phase sulfur has been dissipated or converted.

Using the chemical kinetics described in Section \ref{sec:chem}, as well as the initialisation of the sulfur content of planetesimals described in Section \ref{sec:pfm}, we track the accreted masses of small grains and pebbles, planetesimals, and gas. The tracked mass of the gas-phase sulfur was converted into an atomic abundance with respect to H in the atmospheres of planets that were able to accrete a gas envelope. The accreted sulfur is assumed to be uniform through the gas envelope. Figure \ref{fig:chem_tracks} shows the final sulfur abundances in the cores and atmospheres when only birth time is varied across model runs.

Analysis of our simulated gas giant populations ($M\geq 50 M_{\oplus}$) reveals stark differences in sulfur enrichment across ablation scenarios:

For $p_{\rm ratio}=0$, approximately 61\% of gas giants across all three disk models exceed the Solar S/H value, but only up to a maximum of $1.38$ times. There is little variance in the final bulk atmospheric composition, with all gas giants confined to $0.7-1.38$ times Solar sulfur abundance. This narrow band cannot explain any of the outer Solar System giants, which range between $3-46$ times $\rm S/H_{\odot}$ \citep{briggs1989, wong2004, tollefson2021, molter2021}. Including partial planetesimal ablation ($p_{\rm ratio}=0.5$) drastically changes the previously limited range of sulfur abundances. Similarly to the previous case, approximately 60\% of simulated gas giants across all three disk models exceeded $\rm S/H_{\odot}$, but 10\% exceeded two times this, and 0.13\% exceeded five times $\rm S/H_{\odot}$ and reaching a maximum atmospheric sulfur abundance of $8.37\times10^{-5}$, approximately six times the Solar value. 

In the case of total planetesimal ablation in gas giant atmospheres ($p_{\rm ratio}=1$) 7.56\% of them exceed $3\times \rm S/H_{\odot}$, and 0.16\% exceed eight times the Solar value. We find that including planetesimal ablation in gas giant atmospheres spawns enormous diversity in sulfur abundance, spanning roughly six orders of magnitude in the case of $p_{\rm ratio}=1$. 

\begin{figure*}
\includegraphics[clip=,width=1\linewidth]{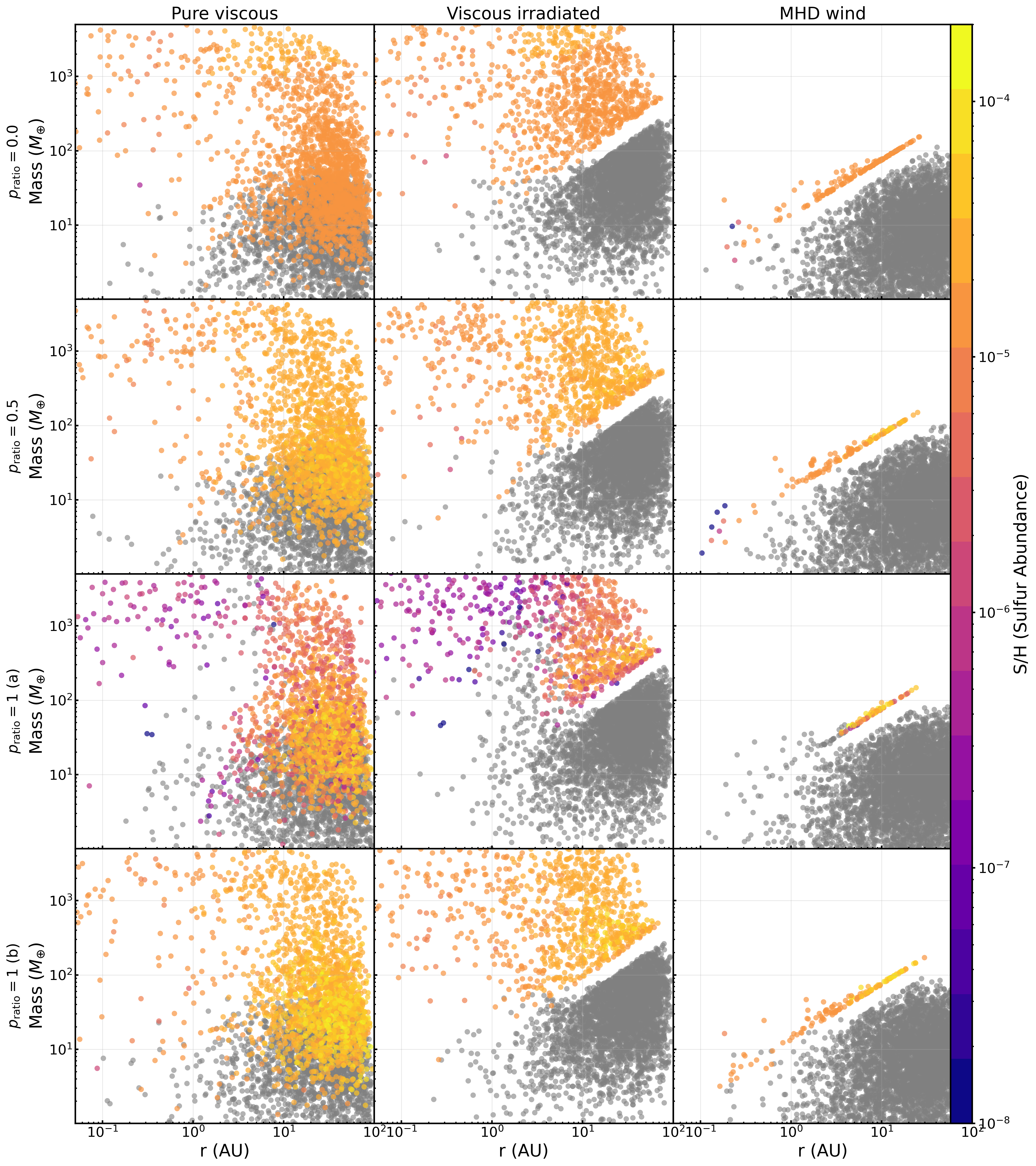}
\caption{Populations generated in all three considered disk models with our three $p_{\rm ratio}$ values. 5000 planets are generated for each case using the parameter ranges given in Table \ref{table:1}. Two cases of $p_{\rm ratio}=1.0$ are shown, one where gas-phase sulfur is omitted in the disk (a), and one where $\mathrm{H_2S}$ gas is present and able to react with iron grains (Section \ref{sec:chem}) (b). Grey points are final planets that have either not undergone any gas accretion, or only possess a negligible/ no amount of sulfur in their envelopes ($S/H \leq10^{-10}$).}
\label{fig:ninecases}
\end{figure*}

The transition from $p_{\rm ratio}=0$ to $p{\rm ratio}=0.5$ reveals a striking distribution in atmospheric sulfur content (Figure \ref{fig:ninecases}). While gas-only accretion produces a single, narrow population clustered around Solar abundance ($0.7-1.38\times$), introducing planetesimal ablation splits the population into two distinct groups: a `normal' population that remains near Solar values and a `enhanced' population representing planets that accreted sulfur-rich planetesimals from beyond the $\rm H_{2}S$ snowline. This trend becomes even more pronounced at $p_{\rm ratio}=1$, where the gap between the two populations widens, with the enhanced' group extending to $9.5\times$ Solar. The persistence of the sulfur-normal population even with complete planetesimal ablation indicates that many gas giants either form in regions already depleted of volatile sulfur or exhaust their local planetesimal reservoirs before significant envelope growth.

\begin{figure} 
\includegraphics[clip=,width=1\linewidth]{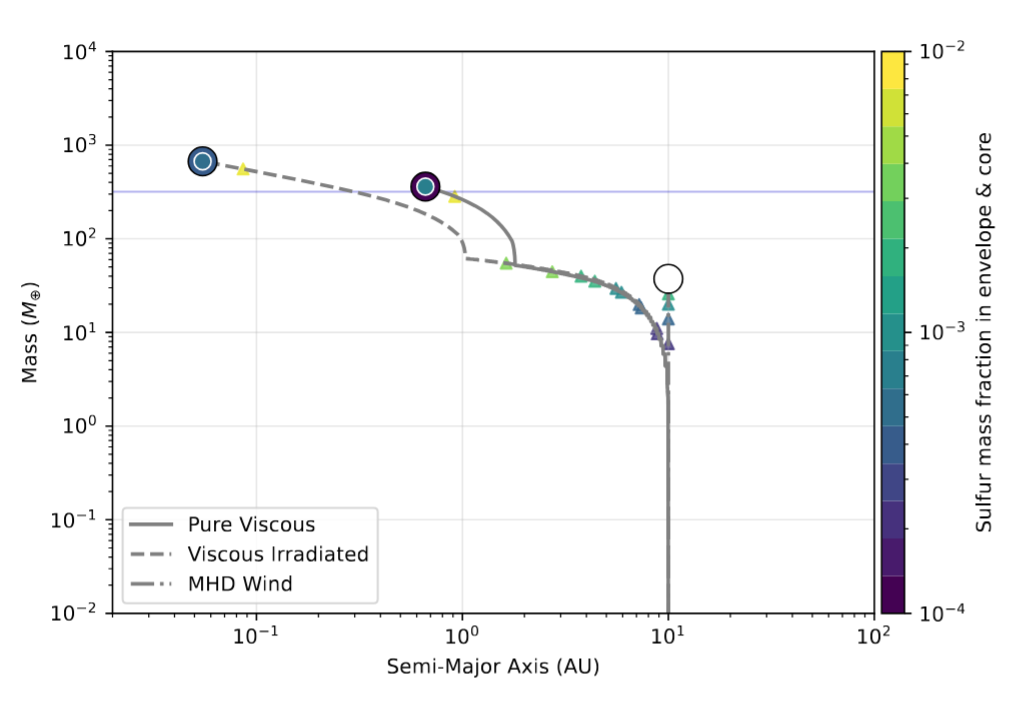}
\caption{Multiple evolution tracks generated from same fiducial disk and planet input parameters, same birth time ($0.1Myrs$) and location (10\,AU), only varying disk type. Line-style denotes disk model, and shown in the figure's legend. Triangles indicate time evolution. Growth tracks are marked every $0.5Myrs$.}
\label{fig:multidisk}
\end{figure}

\begin{figure*}
\includegraphics[clip=,width=1\linewidth]{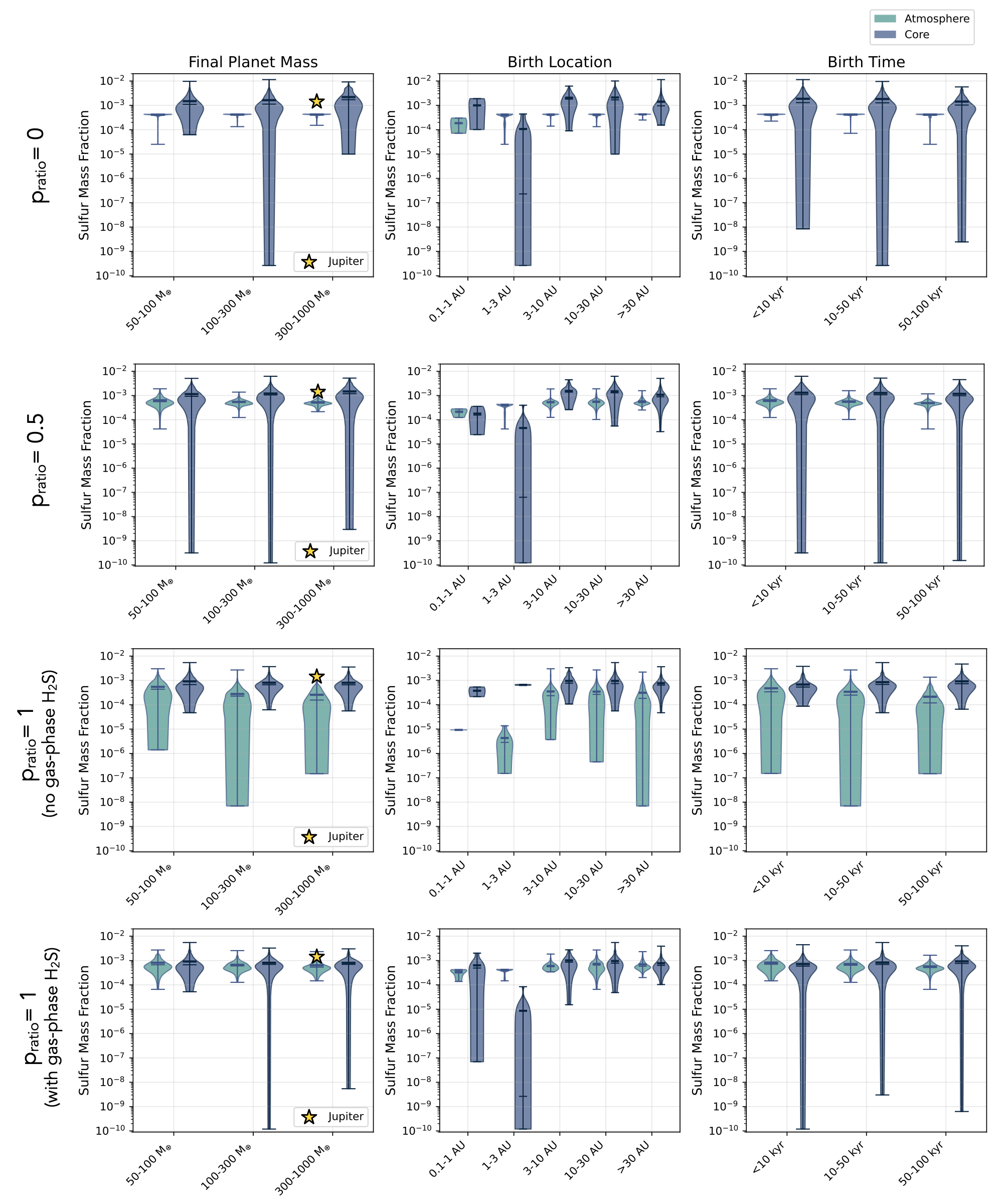}
\caption{Violin plots showing the non-zero sulfur mass fractions in the cores and envelopes of giant planets (0.2 -- 2\, $M_{\rm J}$) across all three disk models. Each row corresponds to one of the three values considered for the ratio of planetesimals ablated in the planets' atmospheres. Two cases of $p_{\rm ratio}=1$ are featured: one where sulfur is only considered in the refractory carried by planetesimals, and one where the chemical kinetics described in Section\,\ref{sec:chem} are included, therefore including both gas-phase and refractory sulfur reservoirs (this is also the model used for $p_{\rm ratio}=0$ and $0.5$)}. Jupiter's atmospheric sulfur mass fraction is marked in its appropriate mass category for reference.
\label{fig:violin}
\end{figure*}

\begin{figure*} 
\includegraphics[clip=,width=1\linewidth]{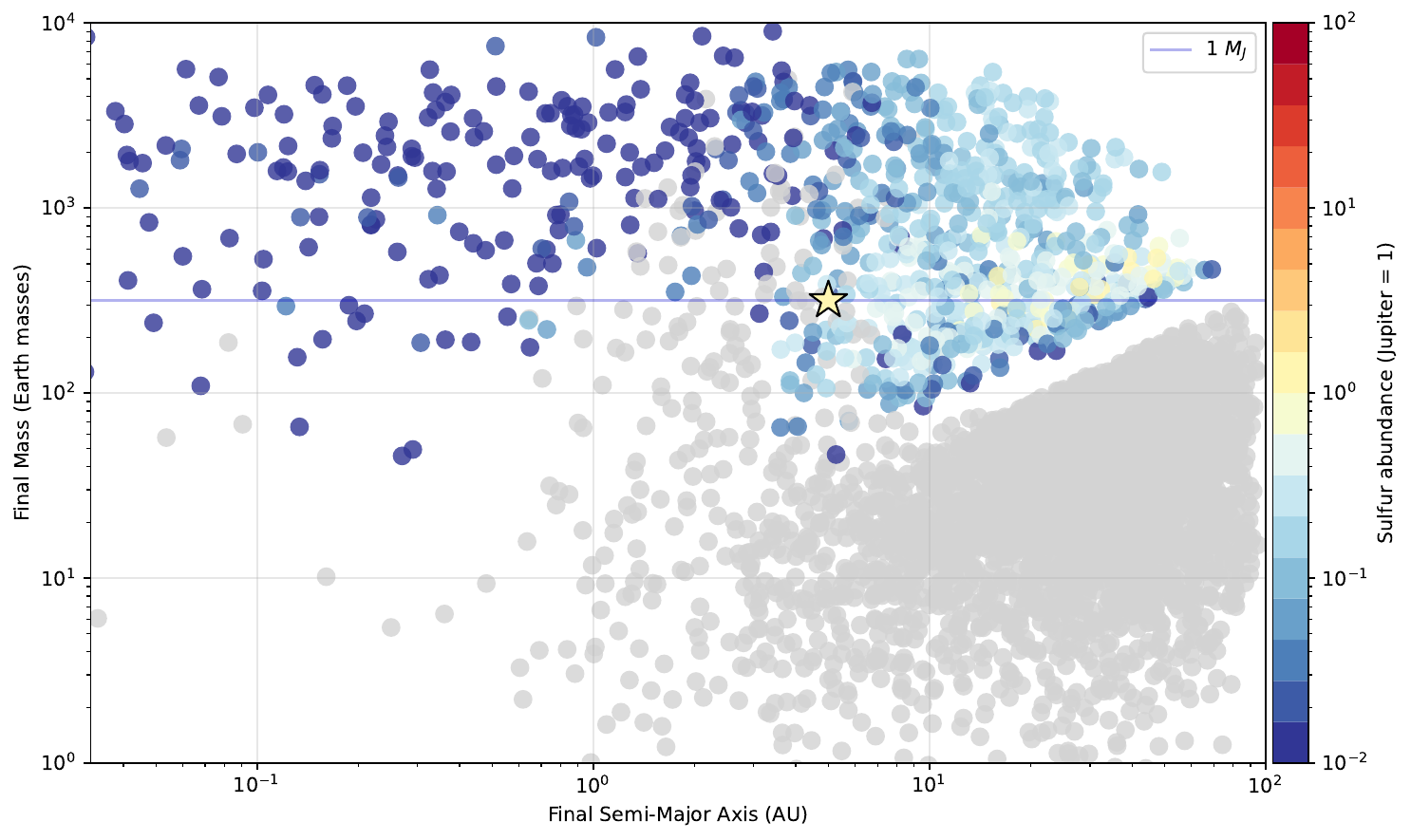}
\caption{Final mass and semi-major axis of population of planets generated using the viscous irradiated disk model, $p_{\rm ratio}= 1$, without gas phase sulfur present. Final envelope sulfur abundance normalised to Jupiter from \protect\cite{wong2004}. Grey points are final planets that have either not undergone any gas accretion, or only possess a negligible/ no amount of sulfur in their envelopes.}
\label{fig:jupe_normal}
\end{figure*}

%\begin{figure*} 
%\includegraphics[clip=,width=1\linewidth]{FIG_sulfurmassvsma.pdf}
%\caption{Final semi-major axis against final envelope sulfur mass fraction. Marker size and colour denote final planet mass and birth location respectively. This is the population generated in the viscous irradiated disk model, with $p_{\rm ratio}=1$, such that all planetesimals contribute all of their mass to the envelope upon accretion, and no sulfur is accreted during runaway gas accretion via $\mathrm{H_2S}$.}
%\label{fig:smassvsma}
%\end{figure*}

\section{Discussion}\label{sec:discussion}

%\mk{Some thoughts:
%\begin{enumerate}
%\item Are your planet formation tracks similar to those from other popsyn codes? Where they significantly differ, what might be the reason(s)?

%\item How would you summarise the range of diversity (or lack thereof in some parts of parameter space) in the sulfur content of planetary cores and envelopes? Are there robust findings such as planetary cores forming outside a certain temperature regime always end up sulfur-poor? What are the possible meanings of finding a high/nominal/low sulfur content in the gas envelope of a giant planet? etc.

%\item What is known about the range of sulfur content in giant planet atmospheres and how does that look in light of your results? Jupiter -- can it only be explained by late-stage planetesimal pollution, or are there other ideas you can put forward? What about the sulfur content of those where JWST has found SO$_{2}$ or H$_{2}$S?

%\item Any significant caveats of the planet formation and sulfur chemistry models?

%\item RESULST range of sulfur. isolate categories: jupiter 0.2 to 1.2 jupes. ice giants mass wise. terrestrials. earth to super earth. sulfr core content. inner vs outer disk etc. for discssions:  where are the sulfur rich giant planets? observations? is that an artifact of the model etc tec? why? why not?
%\end{enumerate}
%}

\subsection{Planet formation outcomes}

Initial disk conditions play a key role in determining the final outcomes of planet formation both physically and chemically. Initial disk mass is directly responsible for the availability of accretable material for planets, as well as the the starting mass of solids in the disk before the dust-to-gas ratio is altered by pebble drift/accretion. Furthermore, alpha viscosity plays a large role in the availability of pebbles, due to its drastic influence on pebble flux. Lower turbulence allows for more efficient dust settling, allowing for a higher Stokes number to be reached, peaking at $St=1$ (Eq. \ref{eq:drift}), which is conducive to efficient pebble drift.

Figure \ref{fig:all_disk_profiles} shows the evolution of midplane temperature and surface density for the three disk models considered in this study, using the fiducial input parameters given in Table \ref{table:1}, and Figure \ref{fig:evotracks_nochem} shows nine evolution tracks using those parameters in the case of the viscous irradiated disk model. When considering other population synthesis/ planet formation models \citep{chambers2018, newgen, chemcomp, simab, savvidou2023} that simulate gas giant formation, the final planets produced by \textsc{sponchpop} vary. 

In regards to simpler planet formation models such as \cite{chambers2018}, \cite{simab}, and \cite{savvidou2023}, \textsc{sponchpop} produces smaller final planets. This discrepancy is caused by the fact that our model removes solid material (both pebble and planetesimal) from the disk as it is accreted by the protoplanet, and pebbles are also drifting inwards at a radial drift velocity dependant on the chosen alpha viscosity parameter. When removing this feedback from the planet to the disk, \textsc{sponchpop}'s planet formation model produces final planets of similar final masses and locations as produced by these models, but only in the case of the viscous-irradiated and pure viscous disk models. The MHD wind model, in comparison, produces an extremely small number of gas giants, as the wind rapidly removes material before it can be accreted by a growing protoplanet.

Higher complexity models, such as \cite{chemcomp} and \cite{newgen}, are able to evolve multiple protoplanets in the same disk and also consider outwards migration, which \textsc{sponchpop} is not yet capable of. Our planet formation model also differs in the fact that a radially dependant initial embryo size is not used. Radially dependant embryo sizes are informed by the mass at which pebble accretion becomes efficient over planetesimal accretion \citep{voelkel2020}. Similarly to \cite{chambers2018}, a constant embryo birth mass is used regardless of local disk conditions, motivated by the fact that both pebble and planetesimal accretion are considered for embryos across the disk. Subsequently, this choice produces negligible difference to the final planets produced by \textsc{sponchpop} when compared.

\subsection{Disk models}
As displayed in Figure \ref{fig:multidisk}, when the same input disk parameters are applied across varying disk models, there is still large disparity in not only the final sulfur abundances in the core and envelope of the planets, but also their final locations and masses. These differences can be attributed to the the distinctions in the dynamics of each disk. In the case of the viscous irradiated disk model, stellar irradiation incident on the outer radii of the disk leads to an increased gas scale height, facilitating a higher pebble isolation mass in the outer disk. Since the disk models used in \textsc{sponchpop} do not consider stellar evolution, and all populations are generated around solar analogues, this means the pebbles' isolation mass in the outer disk stays relatively constant, even with other varying disk parameters. The same applies for the MHD wind disk model, as this model also has stellar irradiation incident on the outer radii.

Statistical robustness varies across the disk models considered. While the viscous and viscous-irradiated models each produce hundreds of gas giants, the MHD wind model generates a relatively small number of gas giants ($\le 50$ for each $p_{\rm ratio}$ case). When MHD results are omitted from our analysis, the final distributions remain indistinguishable, indicating that our giant planet populations are dominated by the other two disk evolution models. The observed exoplanet sample may have contributions from different types of disks, where some types have a higher and others a lower formation likelihood of gas giants.

While the chemical model presented here is a simplification of gas-grain sulfur conversion, $\mathrm{H_2S}$ and FeS represent key reservoirs of volatile and refractory sulfur. We also plan to include a more diverse set of species using more complex chemical networks in the future. 

\subsection{FeS formation model assumptions}

We have assumed an FeS formation rate corresponding to a 1mm grain size. For a fixed mass of dust, this sets the surface area available to convert gas-phase H$_{2}$S into solid FeS. While the FeS formation timescale is sensitive to the solid Fe mass per unit reactive surface area, the spatial extent of the `sulfur desert' is primarily controlled by temperature-dependent sublimation and freeze-out processes rather than grain size variations. Smaller grains present in the ISM ($\sim \mathrm{0.1\,\mu m}$) would increase reactive surface area and potentially expand the sulfur desert, while larger 1\,cm pebbles would have the opposite effect. Compared to the dramatic temperature effects governing sublimation and freeze-out, grain size variation provides less significant changes in FeS spatial distribution.

While the total grain surface area available for the FeS formation reaction does depend on the assumed grain size, and thus so does the exact timescale of full H$_{2}$S to FeS conversion, in reality most of the dust mass will be in large particles. Therefore, we consider it more realistic to approximate the dust mass with $\sim1\,$mm grains rather than more ISM-like $\sim0.1\,\mu$m grains.

The assumption of location and temperature-dependent planetesimal composition initialized at time $t=0$ maximizes the solid `sulfur desert' effect in our models, particularly at higher $p_{\rm ratio}$ values, since planetesimals do not experience the same radial drift as smaller grains and pebbles. This approach reveals the spatial dependence of sulfur partitioning that we aim to investigate. In contrast, assuming a uniform distribution of sulfur-rich planetesimals across the entire disk would mitigate this location-dependent sulfur chemistry effect. Indeed, other planet population synthesis models that treat sulfur as uniformly refractory throughout the disk \citep{turrini2021tracing, pacetti2022} do not reproduce the sulfur desert phenomenon we identify here.

We have furthermore neglected any higher-order effects in the H$_{2}$S--FeS gas-grain system, such as deeper Fe-layers being increasingly masked from surface-level sulfur by the growing FeS layer \citep{lauretta1996rate}. This would lengthen the timescale for all S to be incorporated in FeS, but due to the exponential temperature dependence of the overall reaction rate as well as the potential differences between the laboratory test samples and real protoplanetary grains, we have chosen to neglect these effects in the current study.

\subsection{Final sulfur abundances}

As shown in Figure \ref{fig:ninecases}, the range of final envelope sulfur abundance increases with the increase in our $p_{\rm ratio}$ value. This means the inverse is also true in the case of the sulfur abundance of the planet cores, less sulfur is contributed to the the cores of the planets during late-stage planetesimal accretion.

In the $p_{\rm ratio}=0$ case, the maximum sulfur abundance achieved by a gas giant ($\geq 50M_{\oplus}$) was 1.4 times the present day solar abundance. The maximums reached by the $p_{\rm ratio}=0.5$ and $p_{\rm ratio}=1$ were 5.9 and 9.5 times solar, respectively. The most similar planet to Jupiter generated in our models (with a mass of $321.47 \rm M_{\oplus}$ and final semi-major axis of $5.16 \rm AU$) produced a final envelope sulfur abundance of 1.37 times the solar sulfur abundance. In order to achieve Jupiter's sulfur abundance of $4.45\times10^{-5}$, the planet would need an additional $0.0165M_{\oplus}$ of sulfur in its atmosphere, whether through bombardment of sulfur-rich planetesimals, late stage cometary bombardment, or significant core-envelope mixing. However, Jupiter's enhanced sulfur abundance is able to be met by comparably sized cold gas giant planets in the case of $p_{\rm ratio} = 0.5$ if accreting sulfur-rich planetesimals beyond the $\mathrm{H_2S}$ iceline. This only true of planets that have migrated into regions of the disk that have not been depleted of sulfur rich planetesimals during the core accretion stage. This is dependant on the alpha viscosity of the disk, which heavily influences the migration speed of the planet, particularly during Type\,II migration. 

The most striking result from our simulations is the difference in final sulfur abundance between gas giants depending on where they underwent the majority of their gas accretion phase. The observed sulfur enrichment via late-stage planetesimal infall is made apparent by the difference in maximum sulfur abundances between planets formed in simulations with non-zero $p_{\rm ratio}$ values. For our $p_{\rm ratio}=0$ model runs, the little variance in the final bulk sulfur content from the protosolar S/H is due to the region of the disk dominated by efficient FeS formation being relatively close in to the host star, around 0.4 to 3.5AU in the fiducial cases of the pure viscous and viscously irradiated disk models. Typically, giant planets that finish their formation in this region of the disk did not form in situ, but rather originated further out in more turbulent disks when volatile sulfur was still abundant, and then migrated inwards. It is known that gas giants form most efficiently in the outer disk \citep{lambrechts2012}, and are not likely to begin formation in the inner disk, a region where gas-phase sulfur is depleted by FeS formation.

Similarly, with this increase of planetesimal bombardment (our $p_{\rm ratio}$ value), the inverse is true for the sulfur content in the cores of planets. The evolution of the core sulfur content reveals an equally important but inverse trend. At $p_{\rm ratio}=0$, planetary cores show a strong spatially dependent sulfur mass fractions spanning from $10^{-10}$ to $10^{-2}$, with the bulk of the sulfur depletion occurring in planets born from $1-3\rm AU$. While seemingly a counter-intuitive result,  when considering the location of efficient FeS formation in our fiducial models (Figure \ref{fig:species_evo}), this location in the disk across model runs represents an area between the refractory sulfur reservoirs (the zone of efficient FeS formation and $\mathrm{H_2S}$ ice rich planetesimals) and areas that previously had iron-sulphide rich pebbles before they drifted inwards. This trend becomes more extreme for the $p_{\rm ratio}=0.5$ and $p_{\rm ratio}=1$ cases. This observation--- that increased planetesimal ablation leads to more S-depleted cores--- occurs because planetesimals that would have contributed sulfur to cores instead ablate in the envelope. 

Our results indicate that the atmospheric sulfur content of giant planets is fundamentally tied to late stage accretion processes rather than being dependant on the presence of gas-phase sulfur alone.

Typically, across the populations presented in Figure \ref{fig:ninecases}, high sulfur abundance in gas envelopes can be attributed to formation beyond the $\mathrm{H_2S}$ snowline combined with significant planetesimal ablation during gas accretion. Atmospheric sulfur abundances around the Solar value, as seen primarily in envelopes not supplemented by planetesimal accretion, can be indicative of planets that have accreted gas beyond the FeS stable region, i.e. where $\mathrm{H_2S}$ gas has not been depleted due to efficient FeS formation. The most sulfur poor envelopes belong to planets that spend the majority of their formation histories in the area of the disk when FeS formation has exhausted the local gas-phase sulfur, or formed and remained in the outer disk having accreted all local sulfur-ice-rich planetesimals to their cores before the onset of runaway gas accretion.

This processing of sulfur, dependant on the local physical conditions and initial abundances, lends itself as a possible formation tracer as the atmospheric sulfur abundances directly reflect the thermal environment during the runaway gas accretion phase, extending and complimenting traditional C/O based approaches.

While total and consistent planetesimal ablation is an overestimation, $p_{\rm ratio}$ of $0.5$ may be too low to account for even iron-dominant impactors in Jovian envelopes \citep{pinhas2016}. This ratio would vary with impact velocity, angle, and mass. Ice-dominant planetesimals, as assumed to be the case for the planetesimals formed beyond the $\mathrm{H_2S}$ iceline in this study, would be destroyed upon accretion even more easily due to thermal ablation. Similarly, $p_{\rm ratio} = 0$ is a severe underestimation of planetesimal ablation, even for planets with less substantial gas envelopes, such as Neptune analogues \citep{pinhas2016}.

However, the amount of planetesimals in the outer disk, and the locations they occupy in this model, are also entirely dependant on the outer radius of the disk at $t_{0}$. As our $s_{0}$ value is sampled between 30 and 100AU, our planetesimal disk extends to our maximum value. As the planetesimal region holding $\mathrm{H_2S}$ ices extends from the iceline to the outer edge of the disk, this biases simulations of planets further out towards higher sulfur abundances, as the planet will migrate through a larger span of sulfur rich planetesimals, rather than depleting them locally before significant sulfur enhancement can take place, as is the case for outer planets formed in smaller starting disks.

The core sulfur distributions in Figure \ref{fig:violin} reveal an unexpected and robust depletion pattern that challenges conventional assumptions about refractory sulfur accretion during planet formation. In the case of the cores of giant planets/ super-Earths, there is an evolution from broad, location-dependent sulfur content when planetesimal accretion contributes to the core ($\rm p_{ratio}\leq0.5$, panels a and b of Figure \ref{fig:violin}) to narrower, location-independent sulfur content with ablation entirely in the envelope ($\rm p_{ratio} = 1$, panel c). Without the consideration of volatile sulfur across the disk, a distinct lack of diversity arises in the sulfur abundances of planetary cores. Cores show consistent depletion occurring for planets born at 0.1-3 AU. While this seems counter-intuitive, as this is the region in our fiducial cases where FeS formation is the most efficient, this is instead caused by disks less massive than our fiducial mass being on average cooler, therefore this `FeS zone' occurs closer to the host star. Subsequently, this core sulfur depletion arises from these planets being born in an area of the disk that is typically too cool for efficient FeS formation and too warm for $\mathrm{H_2S}$ to condense into ices. This depletion is less severe for the case of $p_{\rm ratio} =1$ likely because planetesimals bearing refractory sulfur in the inner disk are not subject to the same dynamics as the iron sulfide pebbles and small grains, which drift inwards dependant on the local stokes number. As the planetesimals are larger, and scattering is not considered in this model, the inner reservoir of sulfur-rich planetesimals remains constant until accreted either into the core or into the envelope via ablation. Simultaneously, this is why the envelope sulfur mass fractions of giant planets are lower in the inner disk, as planets will exhaust the sulfur rich planetesimals, meaning the local reservoir is depleted by the time runaway gas accretion starts.

\subsection{Solar System object comparison}

While Jupiter exhibits a notable sulfur enhancement of its atmosphere \citep{wong2004} that is well documented, this sulfur enrichment is far exceeded by the planets beyond its orbit. Saturn's sulfur abundance in its atmosphere due to $\mathrm{H_2S}$ is 10 times that of the present day solar abundance \citep{briggs1989}. Uranus and Neptune's atmospheric sulfur abundances are even higher at $31$ and $46$ times the present day solar abundance respectively \citep{tollefson2021, molter2021}. Evidently, for the outer Solar System, there is a trend amongst the ice giants of increased $\mathrm{H_2S}$ mixing ratios with increased semi major axis. This is a trend replicated in our simulations when accounting for planetesimal ablation ($p_{\rm ratio}\geq0.5$). Yet, only Jupiter's level of atmospheric enhancement is met by the planets we simulate here (Figure \ref{fig:jupe_normal}). Our $p_{\rm ratio}=1$ model--- where gas-phase sulfur is omitted in the disk--- runs produced a wide range of sulfur contents across Jupiter sized planets. However, similarly sized planets that met Jupiter's atmospheric sulfur abundance were only formed further out in the disk ($\geq 10AU$) from Jupiter's location, beyond the $\mathrm{H_2S}$ iceline. Around Jupiter's semi-major axis there is a population of planets with severely sulfur depleted atmospheres, due to the majority of their formation history taking place in the `sulfur desert' between refractory reservoirs. Still, even with the most extreme case of planetesimal ablation considered here, the extremes achieved by the Solar System's ice giants are not replicated in this study. Further sulfur enrichment might have been achieved in the outer Solar System giants by core-envelope mixing caused by the same planetesimal bombardment supplying these planets' envelopes \citep{wahl2017}. 

The extreme sulfur enhancements observed in Uranus ($31\times$ solar) and Neptune ($46\times$ solar) remain beyond the reach of our models, which achieve maximum enrichments of $\sim 9.5 \times$ the solar value even with complete planetesimal ablation. These ice giant abundances likely require additional enrichment mechanisms not captured in our 3 Myr formation time-frame. Cometary bombardment during later evolutionary phases represents a promising avenue for achieving such extreme sulfur abundances, as comets are highly sulfur-rich in both volatile $\mathrm{H_2S}$ and refractory \citep{calmonte2016sulphur}, and would ablate efficiently in ice giant atmospheres. While our population synthesis focuses on early formation processes around solar analogues rather than specifically reproducing Solar System architecture, future work incorporating extended evolutionary timescales and cometary contributions could address this discrepancy.

\section{Conclusions}

In this work, we present the population synthesis model \texttt{sponchpop} and its planet formation module and incorporate a novel gas-grain sulfur chemical model, coupling consideration of both volatile and refractory sulfur in protoplanetary disks. With it, we investigate the information envelope sulfur abundances provide regarding possible formation histories. By tracking the accretion of sulfur through gas-grain conversion and considering varying degrees of late-stage planetesimal infall, we investigate how sulfur abundances in gaseous envelopes can provide information regarding formation histories.

We generated multiple planets at a variety of different birth locations and birth times for disks of varying viscosities, initial masses, and models. Key elemental abundances in the disks considered for this study are assumed to be inherited from their host stars, in this case, solar analogues. 

Our key findings are:
\begin{enumerate}
\item[1.] \textbf{Sulfur content in planet cores and rocky planets is sensitive to a coupled treatment of volatile and refractory sulfur} and is intrinsically related to the formation and migration history. Cores born in the inner few AU ($\leq 3\rm AU$ in our fiducial model) lack refractory sulfur, as the temperature range for efficient FeS formation or retention is conducive to a much smaller region of the disk than previously assumed. Treating sulfur as only being carried in refractories neglects the solid sulfur `desert' which our models predict, where neither efficient iron sulfide production nor $\mathrm{H_2S}$ freeze-out is relevant. This fundamentally changes sulfur core budget predictions for gas giant cores and terrestrial planets. Refractory sulfur in inner disk cores would need a later FeS formation time from replenished H$_{2}$S gas or the inward migration of S-rich planetesimals or cores from further out.
\item[2.] \textbf{Atmospheric sulfur abundances indicate formation and migration histories} for planets with substantial envelopes. Super-solar atmospheric sulfur requires formation beyond the $\mathrm{H_2S}$ iceline as well as efficient planetesimal ablation in the outer disk after the onset of runaway gas accretion. Significant sulfur depletion ($< 10^{-8}$) indicates either formation in volatile sulfur depleted zones due to efficient refractory S conversion, or lack of access to sulfur bearing planetesimals for late-stage infall. This formation diagnostic can compliment standard C/O analysis based approaches to isolate possible migration histories. The dependence of sulfur chemistry on the local thermal and pressure conditions of the disk midplane offers information regarding the formation histories of gas-giants. Migration histories generate core-envelope compositions that indicate possible relative timing of solid and gas accretion phases, as a coupled treatment of volatile and refractory sulfur can provide information regarding the timing of solid and gas accretion through untouched or depleted planetesimal reservoirs.
\item[3.] \textbf{Jupiter and other Solar System giants require efficient late-stage planetesimal ablation} to reach observed enhanced sulfur abundances. The maximum envelope abundance achieved without planetesimal ablation is $\mathrm{S/H}=1.95\times10^{-5}$, and while slightly higher ($\times 1.4$) than the present day solar sulfur abundance, still falls short of Jupiter's sulfur abundance of $\mathrm{S/H}= 4.45\times10^{-5}$ by a factor of $\sim2-3$. This indicates that efficient planetesimal ablation or core-envelope mixing is necessary for reproducing the trend seen in the outer planets of the Solar System to supplement any gas-phase sulfur accreted. Super-solar abundances are only consistently produced when the fraction of accreted planetesimal mass ablated into the envelope ($p_{\rm ratio}$) is high enough, $p_{\rm ratio} \geq 0.5$, suggesting that all of the Solar System giants experienced significant late stage planetesimal bombardment that enhanced their bulk atmospheric sulfur content.
\end{enumerate}

Our results establish sulfur chemistry as a more powerful tool for formation pathway reconstruction than had previously been realised, while highlighting the complex interplay between disk chemistry, formation timescales, and migration history in determining the final composition of planetary cores and atmospheres. The framework presented here provides a foundation for incorporating realistic volatile sulfur chemistry into planet formation and population synthesis models, essential for fully exploiting the formation information encoded in exoplanet atmospheric observations. The presence of sulfur in the atmospheres of gas giants can be attributed to the formation histories of these planets, particularly in the outer disk. While the methods presented here do not account for other possible theories explaining the presence of volatile sulfur in gas-giant atmospheres such as core-envelope mixing, sulfur's abundance in gas giant atmospheres remain sensitive to formation history of these bodies.

As JWST and future facilities such as \emph{Ariel} expand the sample of planets with measured atmospheric sulfur abundances, population synthesis approaches incorporating detailed chemical evolution will become increasingly crucial for linking protoplanetary disk observations to mature planetary systems, advancing our understanding of the formation processes that produced the diverse exoplanet population observed today.

\section*{Acknowledgements}

This project has received funding from the European Union's Horizon Europe research and innovation program under grant agreement No.\,101079231 (EXOHOST) and from the United Kingdom Research and Innovation (UKRI) Horizon Europe Guarantee Scheme with grant No.\,10051045. OS acknowledges support from UKRI grant UKRI1184. We thank Sebastiaan Krijt, Paola Pinilla, Luke Keyte and Joe Williams for their insightful feedback. We would also like to thank the anonymous referee for their helpful suggestions and feedback.

%%%%%%%%%%%%%%%%%%%%%%%%%%%%%%%%%%%%%%%%%%%%%%%%%%
\section*{Data Availability}
The data underlying this article will be shared on reasonable request to the corresponding author.

%The inclusion of a Data Availability Statement is a requirement for articles published in MNRAS. Data Availability Statements provide a standardised format for readers to understand the availability of data underlying the research results described in the article. The statement may refer to original data generated in the course of the study or to third-party data analysed in the article. The statement should describe and provide means of access, where possible, by linking to the data or providing the required accession numbers for the relevant databases or DOIs.


\bibliographystyle{mnras}
\bibliography{p1}

@article{safronov1969,
  title={Relative sizes of the largest bodies during the accumulation of planets},
  author={Safronov, VS},
  journal={Icarus},
  volume={10},
  number={1},
  pages={109--115},
  year={1969},
  publisher={Elsevier}
}

@article{mordasini2009,
  title={Extrasolar planet population synthesis-I. Method, formation tracks, and mass-distance distribution},
  author={Mordasini, Christoph and Alibert, Yann and Benz, Willy},
  journal={Astronomy \& Astrophysics},
  volume={501},
  number={3},
  pages={1139--1160},
  year={2009},
  publisher={EDP Sciences}
}

@article{chambers2018,
  title={Planet formation: an optimized population-synthesis approach},
  author={Chambers, John},
  journal={The Astrophysical Journal},
  volume={865},
  number={1},
  pages={30},
  year={2018},
  publisher={IOP Publishing}
}

@article{newgen,
  title={The New Generation Planetary Population Synthesis (NGPPS)-I. Bern global model of planet formation and evolution, model tests, and emerging planetary systems},
  author={Emsenhuber, Alexandre and Mordasini, Christoph and Burn, Remo and Alibert, Yann and Benz, Willy and Asphaug, Erik},
  journal={Astronomy \& Astrophysics},
  volume={656},
  pages={A69},
  year={2021},
  publisher={EDP Sciences}
}

@article{simab,
  title={SimAb: A simple, fast, and flexible model to assess the effects of planet formation on the atmospheric composition of gas giants},
  author={Khorshid, N and Min, M and D{\'e}sert, JM and Woitke, P and Dominik, C},
  journal={Astronomy \& Astrophysics},
  volume={667},
  pages={A147},
  year={2022},
  publisher={EDP Sciences}
}

@article{piso2014,
  title={On the minimum core mass for giant planet formation at wide separations},
  author={Piso, Ana-Maria A and Youdin, Andrew N},
  journal={The Astrophysical Journal},
  volume={786},
  number={1},
  pages={21},
  year={2014},
  publisher={IOP Publishing}
}

@article{machida2010,
  title={Gas accretion onto a protoplanet and formation of a gas giant planet},
  author={Machida, Masahiro N and Kokubo, Eiichiro and Inutsuka, Shu-ichiro and Matsumoto, Tomoaki},
  journal={Monthly Notices of the Royal Astronomical Society},
  volume={405},
  number={2},
  pages={1227--1243},
  year={2010},
  publisher={Blackwell Publishing Ltd Oxford, UK}
}

@article{tanaka2002,
  title={Three-dimensional interaction between a planet and an isothermal gaseous disk. I. Corotation and Lindblad torques and planet migration},
  author={Tanaka, Hidekazu and Takeuchi, Taku and Ward, William R},
  journal={The Astrophysical Journal},
  volume={565},
  number={2},
  pages={1257},
  year={2002},
  publisher={IOP Publishing}
}

@article{bitsch2019,
  title={Formation of planetary systems by pebble accretion and migration: growth of gas giants},
  author={Bitsch, Bertram and Izidoro, Andre and Johansen, Anders and Raymond, Sean N and Morbidelli, Alessandro and Lambrechts, Michiel and Jacobson, Seth A},
  journal={Astronomy \& Astrophysics},
  volume={623},
  pages={A88},
  year={2019},
  publisher={EDP Sciences}
}

@article{scardoni2020,
  title={Type II migration strikes back--an old paradigm for planet migration in discs},
  author={Scardoni, Chiara E and Rosotti, Giovanni P and Lodato, Giuseppe and Clarke, Cathie J},
  journal={Monthly Notices of the Royal Astronomical Society},
  volume={492},
  number={1},
  pages={1318--1328},
  year={2020},
  publisher={Oxford University Press}
}

@article{pebpre,
  title={How dust fragmentation may be beneficial to planetary growth by pebble accretion},
  author={Dr{\k{a}}{\.z}kowska, J and Stammler, Sebastian M and Birnstiel, Til},
  journal={Astronomy \& Astrophysics},
  volume={647},
  pages={A15},
  year={2021},
  publisher={EDP Sciences}
}

@article{twopoppy,
  title={A simple model for the evolution of the dust population in protoplanetary disks},
  author={Birnstiel, T and Klahr, H and Ercolano, B},
  journal={Astronomy \& Astrophysics},
  volume={539},
  pages={A148},
  year={2012},
  publisher={EDP Sciences}
}

@article{thommes2003,
  title={Oligarchic growth of giant planets},
  author={Thommes, Edward W and Duncan, Martin J and Levison, Harold F},
  journal={Icarus},
  volume={161},
  number={2},
  pages={431--455},
  year={2003},
  publisher={Elsevier}
}

@article{ormel2010,
  title={The effect of gas drag on the growth of protoplanets-analytical expressions for the accretion of small bodies in laminar disks},
  author={Ormel, CW and Klahr, HH},
  journal={Astronomy \& Astrophysics},
  volume={520},
  pages={A43},
  year={2010},
  publisher={EDP Sciences}
}

@article{w1997,
  title={Aerodynamics of solid bodies in the solar nebula},
  author={Weidenschilling, SJ},
  journal={Monthly Notices of the Royal Astronomical Society},
  volume={180},
  number={2},
  pages={57--70},
  year={1977},
  publisher={The Royal Astronomical Society}
}

@article{lambrechts2014,
  title={Separating gas-giant and ice-giant planets by halting pebble accretion},
  author={Lambrechts, Michiel and Johansen, Anders and Morbidelli, Alessandro},
  journal={Astronomy \& Astrophysics},
  volume={572},
  pages={A35},
  year={2014},
  publisher={EDP Sciences}
}

@article{alibert2005,
  title={Models of giant planet formation with migration and disc evolution},
  author={Alibert, Yann and Mordasini, Christoph and Benz, Willy and Winisdoerffer, Christophe},
  journal={Astronomy \& Astrophysics},
  volume={434},
  number={1},
  pages={343--353},
  year={2005},
  publisher={EDP Sciences}
}

@article{chambers2009,
  title={An analytic model for the evolution of a viscous, irradiated disk},
  author={Chambers, JE},
  journal={The Astrophysical Journal},
  volume={705},
  number={2},
  pages={1206},
  year={2009},
  publisher={IOP Publishing}
}

@article{oberg2011,
  title={The effects of snowlines on C/O in planetary atmospheres},
  author={{\"O}berg, Karin I and Murray-Clay, Ruth and Bergin, Edwin A},
  journal={The Astrophysical Journal Letters},
  volume={743},
  number={1},
  pages={L16},
  year={2011},
  publisher={IOP Publishing}
}

@article{kama2019,
  title={Abundant refractory sulfur in protoplanetary disks},
  author={Kama, Mihkel and Shorttle, Oliver and Jermyn, Adam S and Folsom, Colin P and Furuya, Kenji and Bergin, Edwin A and Walsh, Catherine and Keller, Lindsay},
  journal={The Astrophysical Journal},
  volume={885},
  number={2},
  pages={114},
  year={2019},
  publisher={IOP Publishing}
}

@article{oberg2019jupiter,
  title={Jupiter's composition suggests its core assembled exterior to the N2 snowline},
  author={{\"O}berg, Karin I and Wordsworth, Robin},
  journal={The Astronomical Journal},
  volume={158},
  number={5},
  pages={194},
  year={2019},
  publisher={IOP Publishing}
}

@article{turrini2021tracing,
  title={Tracing the formation history of giant planets in protoplanetary disks with carbon, oxygen, nitrogen, and sulfur},
  author={Turrini, Diego and Schisano, Eugenio and Fonte, Sergio and Molinari, Sergio and Politi, Romolo and Fedele, Davide and Pani{\'c}, O and Kama, Mihkel and Changeat, Quentin and Tinetti, Giovanna},
  journal={The Astrophysical Journal},
  volume={909},
  number={1},
  pages={40},
  year={2021},
  publisher={IOP Publishing}
}

@ARTICLE{walsh2025,
       author = {{Walsh}, Catherine},
        title = "{Linking planet formation to exoplanet characteristics: C/O as a diagnostic of planet formation}",
      journal = {arXiv e-prints},
         year = 2025,
        month = aug,
          eid = {arXiv:2508.09587},
        pages = {arXiv:2508.09587},
          doi = {10.48550/arXiv.2508.09587},
archivePrefix = {arXiv},
       eprint = {2508.09587},
 primaryClass = {astro-ph.EP},
       adsurl = {https://ui.adsabs.harvard.edu/abs/2025arXiv250809587W}
}

@INPROCEEDINGS{tinetti2022,
       author = {{Tinetti}, Giovanna and {Eccleston}, Paul and {Lueftinger}, Theresa and {Salvignol}, Jean-Christophe and {Fahmy}, Salma and {Alves de Oliveira}, Caterina},
        title = "{Ariel: Enabling planetary science across light-years}",
    booktitle = {European Planetary Science Congress},
         year = 2022,
        month = sep,
          eid = {EPSC2022-1114},
        pages = {EPSC2022-1114},
          doi = {10.5194/epsc2022-1114},
archivePrefix = {arXiv},
       eprint = {2104.04824},
 primaryClass = {astro-ph.IM},
       adsurl = {https://ui.adsabs.harvard.edu/abs/2022EPSC...16.1114T}
}

@article{wasson1988compositions,
  title={Compositions of chondrites},
  author={Wasson, John T and Kallemeyn, Gregory W},
  journal={Philosophical Transactions of the Royal Society of London. Series A, Mathematical and Physical Sciences},
  volume={325},
  number={1587},
  pages={535--544},
  year={1988},
  publisher={The Royal Society London}
}

@article{calmonte2016sulphur,
  title={Sulphur-bearing species in the coma of comet 67P/Churyumov--Gerasimenko},
  author={Calmonte, Ursina and Altwegg, Kathrin and Balsiger, Hans and Berthelier, Jean-Jacques and Bieler, Andr{\'e} and Cessateur, G and Dhooghe, F and Van Dishoeck, EF and Fiethe, B and Fuselier, SA and others},
  journal={Monthly Notices of the Royal Astronomical Society},
  volume={462},
  number={Suppl\_1},
  pages={S253--S273},
  year={2016},
  publisher={The Royal Astronomical Society}
}

@article{lauretta1996rate,
  title={The rate of iron sulfide formation in the solar nebula},
  author={Lauretta, Dante S and Kremser, Daniel T and Fegley Jr, Bruce},
  journal={Icarus},
  volume={122},
  number={2},
  pages={288--315},
  year={1996},
  publisher={Elsevier}
}

@ARTICLE{Wakelametal2017,
       author = {{Wakelam}, V. and {Loison}, J. -C. and {Mereau}, R. and {Ruaud}, M.},
        title = "{Binding energies: New values and impact on the efficiency of chemical desorption}",
      journal = {Molecular Astrophysics},
         year = 2017,
        month = mar,
       volume = {6},
        pages = {22-35},
          doi = {10.1016/j.molap.2017.01.002},
archivePrefix = {arXiv},
       eprint = {1701.06492},
 primaryClass = {astro-ph.GA},
       adsurl = {https://ui.adsabs.harvard.edu/abs/2017MolAs...6...22W}
}

@article{jenkins2009,
  title={A unified representation of gas-phase element depletions in the interstellar medium},
  author={Jenkins, Edward B},
  journal={The Astrophysical Journal},
  volume={700},
  number={2},
  pages={1299},
  year={2009},
  publisher={IOP Publishing}
}

@article{asplund2009,
  title={The chemical composition of the Sun},
  author={Asplund, Martin and Grevesse, Nicolas and Sauval, A Jacques and Scott, Pat},
  journal={Annual review of astronomy and astrophysics},
  volume={47},
  number={2009},
  pages={481--522},
  year={2009},
  publisher={Annual Reviews}
}

@article{bolton2017,
  title={Jupiter’s interior and deep atmosphere: The initial pole-to-pole passes with the Juno spacecraft},
  author={Bolton, Scott J and Adriani, Alberto and Adumitroaie, V and Allison, M and Anderson, J and Atreya, S and Bloxham, J and Brown, S and Connerney, JEP and DeJong, E and others},
  journal={Science},
  volume={356},
  number={6340},
  pages={821--825},
  year={2017},
  publisher={American Association for the Advancement of Science}
}

@article{fu2024,
  title={Hydrogen sulfide and metal-enriched atmosphere for a Jupiter-mass exoplanet},
  author={Fu, Guangwei and Welbanks, Luis and Deming, Drake and Inglis, Julie and Zhang, Michael and Lothringer, Joshua and Ih, Jegug and Moses, Julianne I and Schlawin, Everett and Knutson, Heather A and others},
  journal={Nature},
  volume={632},
  number={8026},
  pages={752--756},
  year={2024},
  publisher={Nature Publishing Group UK London}
}

@article{keyte2024,
  title={Spatially resolving the volatile sulfur abundance in the HD 100546 protoplanetary disc},
  author={Keyte, Luke and Kama, Mihkel and Chuang, Ko-Ju and Cleeves, L Ilsedore and Drozdovskaya, Maria N and Furuya, Kenji and Rawlings, Jonathan and Shorttle, Oliver},
  journal={Monthly Notices of the Royal Astronomical Society},
  volume={528},
  number={1},
  pages={388--407},
  year={2024},
  publisher={Oxford University Press}
}

@article{semenov2018,
  title={Chemistry in disks-XI. Sulfur-bearing species as tracers of protoplanetary disk physics and chemistry: the DM Tau case},
  author={Semenov, D and Favre, C and Fedele, D and Guilloteau, S and Teague, R and Henning, Th and Dutrey, Anne and Chapillon, E and Hersant, F and Pi{\'e}tu, V},
  journal={Astronomy \& Astrophysics},
  volume={617},
  pages={A28},
  year={2018},
  publisher={EDP Sciences}
}

@article{dyrek2024,
  title={SO2, silicate clouds, but no CH4 detected in a warm Neptune},
  author={Dyrek, Achr{\`e}ne and Min, Michiel and Decin, Leen and Bouwman, Jeroen and Crouzet, Nicolas and Molli{\`e}re, Paul and Lagage, Pierre-Olivier and Konings, Thomas and Tremblin, Pascal and G{\"u}del, Manuel and others},
  journal={Nature},
  volume={625},
  number={7993},
  pages={51--54},
  year={2024},
  publisher={Nature Publishing Group UK London}
}

@article{alderson2023,
  title={Early Release Science of the exoplanet WASP-39b with JWST NIRSpec G395H},
  author={Alderson, Lili and Wakeford, Hannah R and Alam, Munazza K and Batalha, Natasha E and Lothringer, Joshua D and Adams Redai, Jea and Barat, Saugata and Brande, Jonathan and Damiano, Mario and Daylan, Tansu and others},
  journal={Nature},
  volume={614},
  number={7949},
  pages={664--669},
  year={2023},
  publisher={Nature Publishing Group UK London}
}

@article{booth2023sulphur,
  title={Sulphur monoxide emission tracing an embedded planet in the HD 100546 protoplanetary disk},
  author={Booth, Alice S and Ilee, John D and Walsh, Catherine and Kama, Mihkel and Keyte, Luke and van Dishoeck, Ewine F and Nomura, Hideko},
  journal={Astronomy \& Astrophysics},
  volume={669},
  pages={A53},
  year={2023},
  publisher={EDP Sciences}
}

@article{phuong2018,
  title={First detection of H2S in a protoplanetary disk-The dense GG Tauri A ring},
  author={Phuong, NT and Chapillon, E and Majumdar, L and Dutrey, Anne and Guilloteau, S and Pi{\'e}tu, V and Wakelam, Valentine and Diep, PN and Tang, Y-W and Beck, T and others},
  journal={Astronomy \& Astrophysics},
  volume={616},
  pages={L5},
  year={2018},
  publisher={EDP Sciences}
}

@article{tsai2024,
  title={Biogenic sulfur gases as biosignatures on temperate sub-Neptune waterworlds},
  author={Tsai, Shang-Min and Innes, Hamish and Wogan, Nicholas F and Schwieterman, Edward W},
  journal={The Astrophysical Journal Letters},
  volume={966},
  number={2},
  pages={L24},
  year={2024},
  publisher={IOP Publishing}
}

@article{powell2024,
  title={Sulfur dioxide in the mid-infrared transmission spectrum of WASP-39b},
  author={Powell, Diana and Feinstein, Adina D and Lee, Elspeth KH and Zhang, Michael and Tsai, Shang-Min and Taylor, Jake and Kirk, James and Bell, Taylor and Barstow, Joanna K and Gao, Peter and others},
  journal={Nature},
  volume={626},
  number={8001},
  pages={979--983},
  year={2024},
  publisher={Nature Publishing Group UK London}
}

@article{pacetti2022,
  title={Chemical diversity in protoplanetary disks and its impact on the formation history of giant planets},
  author={Pacetti, Elenia and Turrini, Diego and Schisano, Eugenio and Molinari, Sergio and Fonte, Sergio and Politi, Romolo and Hennebelle, Patrick and Klessen, Ralf and Testi, Leonardo and Lebreuilly, Ugo},
  journal={The Astrophysical Journal},
  volume={937},
  number={1},
  pages={36},
  year={2022},
  publisher={IOP Publishing}
}

@article{leroy2015,
  title={Inventory of the volatiles on comet 67P/Churyumov-Gerasimenko from Rosetta/ROSINA},
  author={Le Roy, L{\'e}na and Altwegg, Kathrin and Balsiger, Hans and Berthelier, Jean-Jacques and Bieler, Andre and Briois, Christelle and Calmonte, Ursina and Combi, Michael R and De Keyser, Johan and Dhooghe, Frederik and others},
  journal={Astronomy \& Astrophysics},
  volume={583},
  pages={A1},
  year={2015},
  publisher={EDP Sciences}
}

@article{savvidou2023,
  title={How to make giant planets via pebble accretion},
  author={Savvidou, Sofia and Bitsch, Bertram},
  journal={Astronomy \& Astrophysics},
  volume={679},
  pages={A42},
  year={2023},
  publisher={EDP Sciences}
}

@article{chemcomp,
  title={How drifting and evaporating pebbles shape giant planets-I. Heavy element content and atmospheric C/O},
  author={Schneider, Aaron David and Bitsch, Bertram},
  journal={Astronomy \& Astrophysics},
  volume={654},
  pages={A71},
  year={2021},
  publisher={EDP Sciences}
}

@article{schafer2017,
  title={Initial mass function of planetesimals formed by the streaming instability},
  author={Sch{\"a}fer, Urs and Yang, Chao-Chin and Johansen, Anders},
  journal={Astronomy \& Astrophysics},
  volume={597},
  pages={A69},
  year={2017},
  publisher={EDP Sciences}
}

@article{liu2019,
  title={Growth after the streaming instability-From planetesimal accretion to pebble accretion},
  author={Liu, Beibei and Ormel, Chris W and Johansen, Anders},
  journal={Astronomy \& Astrophysics},
  volume={624},
  pages={A114},
  year={2019},
  publisher={EDP Sciences}
}

@article{wong2004,
  title={Updated Galileo probe mass spectrometer measurements of carbon, oxygen, nitrogen, and sulfur on Jupiter},
  author={Wong, Michael H and Mahaffy, Paul R and Atreya, Sushil K and Niemann, Hasso B and Owen, Tobias C},
  journal={Icarus},
  volume={171},
  number={1},
  pages={153--170},
  year={2004},
  publisher={Elsevier}
}

@incollection{ormel2017,
  title={The emerging paradigm of pebble accretion},
  author={Ormel, Chris W},
  booktitle={Formation, Evolution, and Dynamics of Young Solar Systems},
  pages={197--228},
  year={2017},
  publisher={Springer}
}

@article{lodders2003,
  title={Solar system abundances and condensation temperatures of the elements},
  author={Lodders, Katharina},
  journal={The Astrophysical Journal},
  volume={591},
  number={2},
  pages={1220},
  year={2003},
  publisher={IOP Publishing}
}

@article{pinhas2016,
  title={Efficiency of planetesimal ablation in giant planetary envelopes},
  author={Pinhas, Arazi and Madhusudhan, Nikku and Clarke, Cathie},
  journal={Monthly Notices of the Royal Astronomical Society},
  volume={463},
  number={4},
  pages={4516--4532},
  year={2016},
  publisher={The Royal Astronomical Society}
}

@article{hobbs2021,
  title={Sulfur chemistry in the atmospheres of warm and hot Jupiters},
  author={Hobbs, R and Rimmer, P B and Shorttle, O and Madhusudhan, N},
  journal={Monthly Notices of the Royal Astronomical Society},
  volume={506},
  number={3},
  pages={3186--3204},
  year={2021},
  publisher={Oxford University Press}
}

@article{gansicke2012,
  title={The chemical diversity of exo-terrestrial planetary debris around white dwarfs},
  author={G{\"a}nsicke, BT and Koester, Detlev and Farihi, J and Girven, Jonathan and Parsons, SG and Breedt, E},
  journal={Monthly Notices of the Royal Astronomical Society},
  volume={424},
  number={1},
  pages={333--347},
  year={2012},
  publisher={The Royal Astronomical Society}
}

@ARTICLE{Xuetal2013,
       author = {{Xu}, S. and {Jura}, M. and {Klein}, B. and {Koester}, D. and {Zuckerman}, B.},
        title = "{Two Beyond-primitive Extrasolar Planetesimals}",
      journal = {\apj},
         year = 2013,
        month = apr,
       volume = {766},
       number = {2},
          eid = {132},
        pages = {132},
          doi = {10.1088/0004-637X/766/2/132},
archivePrefix = {arXiv},
       eprint = {1302.4799},
 primaryClass = {astro-ph.EP},
       adsurl = {https://ui.adsabs.harvard.edu/abs/2013ApJ...766..132X}
}

@article{rosotti2023,
  title={Empirical constraints on turbulence in proto-planetary discs},
  author={Rosotti, Giovanni P},
  journal={New Astronomy Reviews},
  volume={96},
  pages={101674},
  year={2023},
  publisher={Elsevier}
}

@article{benz2014,
  title={Planet population synthesis},
  author={Benz, Willy and Ida, Shigeru and Alibert, Yann and Lin, DNC and Mordasini, Christoph},
  journal={arXiv preprint arXiv:1402.7086},
  year={2014}
}

@article{venturini2016,
  title={Planet formation with envelope enrichment: new insights on planetary diversity},
  author={Venturini, Julia and Alibert, Yann and Benz, Willy},
  journal={Astronomy \& Astrophysics},
  volume={596},
  pages={A90},
  year={2016},
  publisher={EDP Sciences}
}

@article{tong2025,
  title={Turbulence and dust fragility in protoplanetary discs},
  author={Tong, Simin and Alexander, Richard and Rosotti, Giovanni},
  journal={arXiv preprint arXiv:2509.24818},
  year={2025}
}

@article{jiang2024,
  title={Grain-size measurements in protoplanetary disks indicate fragile pebbles and low turbulence},
  author={Jiang, Haochang and Mac{\'\i}as, Enrique and Guerra-Alvarado, Osmar M and Carrasco-Gonz{\'a}lez, Carlos},
  journal={Astronomy \& Astrophysics},
  volume={682},
  pages={A32},
  year={2024},
  publisher={EDP Sciences}
}

@article{drazkowska2018,
  title={Planetesimal formation during protoplanetary disk buildup},
  author={Dr{\k{a}}{\.z}kowska, J and Dullemond, Cornelis P},
  journal={Astronomy \& Astrophysics},
  volume={614},
  pages={A62},
  year={2018},
  publisher={EDP Sciences}
}

@article{briggs1989,
  title={Radio observations of Saturn as a probe of its atmosphere and cloud structure},
  author={Briggs, Franklin H and Sackett, Penny D},
  journal={Icarus},
  volume={80},
  number={1},
  pages={77--103},
  year={1989},
  publisher={Elsevier}
}

@article{tollefson2021,
  title={Neptune's spatial brightness temperature variations from the VLA and Alma},
  author={Tollefson, Joshua and De Pater, Imke and Molter, Edward M and Sault, Robert J and Butler, Bryan J and Luszcz-Cook, Statia and DeBoer, David},
  journal={The Planetary Science Journal},
  volume={2},
  number={3},
  pages={105},
  year={2021},
  publisher={IOP Publishing}
}

@article{molter2021,
  title={Tropospheric composition and circulation of Uranus with ALMA and the VLA},
  author={Molter, Edward M and De Pater, Imke and Luszcz-Cook, Statia and Tollefson, Joshua and Sault, Robert J and Butler, Bryan and De Boer, David},
  journal={The Planetary Science Journal},
  volume={2},
  number={1},
  pages={3},
  year={2021},
  publisher={IOP Publishing}
}

@article{lambrechts2012,
  title={Rapid growth of gas-giant cores by pebble accretion},
  author={Lambrechts, Michiel and Johansen, Anders},
  journal={Astronomy \& Astrophysics},
  volume={544},
  pages={A32},
  year={2012},
  publisher={EDP Sciences}
}

@article{voelkel2020,
  title={Effect of pebble flux-regulated planetesimal formation on giant planet formation},
  author={Voelkel, Oliver and Klahr, Hubert and Mordasini, Christoph and Emsenhuber, Alexandre and Lenz, Christian},
  journal={Astronomy \& Astrophysics},
  volume={642},
  pages={A75},
  year={2020},
  publisher={EDP Sciences}
}

@article{wahl2017,
  title={Comparing Jupiter interior structure models to Juno gravity measurements and the role of a dilute core},
  author={Wahl, Sean M and Hubbard, William B and Militzer, Burkhard and Guillot, Tristan and Miguel, Yamila and Movshovitz, N and Kaspi, Yohai and Helled, R and Reese, D and Galanti, Eli and others},
  journal={Geophysical Research Letters},
  volume={44},
  number={10},
  pages={4649--4659},
  year={2017},
  publisher={Wiley Online Library}
}

@article{fuente2010,
  title={Molecular content of the circumstellar disk in AB Aurigae-First detection of SO in a circumstellar disk},
  author={Fuente, A and Cernicharo, J and Ag{\'u}ndez, M and Bern{\'e}, O and Goicoechea, JR and Alonso-Albi, T and Marcelino, N},
  journal={Astronomy \& Astrophysics},
  volume={524},
  pages={A19},
  year={2010},
  publisher={EDP Sciences}
}

@article{dutrey2011,
  title={Chemistry in disks-V. Sulfur-bearing molecules in the protoplanetary disks surrounding LkCa15, MWC480, DM Tauri, and GO Tauri},
  author={Dutrey, Anne and Wakelam, Valentine and Boehler, Y and Guilloteau, S and Hersant, F and Semenov, Dmitry and Chapillon, E and Henning, Thomas and Pi{\'e}tu, Vincent and Launhardt, Ralf and others},
  journal={Astronomy \& Astrophysics},
  volume={535},
  pages={A104},
  year={2011},
  publisher={EDP Sciences}
}

@article{altwegg2016,
  title={Prebiotic chemicals—amino acid and phosphorus—in the coma of comet 67P/Churyumov-Gerasimenko},
  author={Altwegg, Kathrin and Balsiger, Hans and Bar-Nun, Akiva and Berthelier, Jean-Jacques and Bieler, Andre and Bochsler, Peter and Briois, Christelle and Calmonte, Ursina and Combi, Michael R and Cottin, Herv{\'e} and others},
  journal={Science advances},
  volume={2},
  number={5},
  pages={e1600285},
  year={2016},
  publisher={American Association for the Advancement of Science}
}

%%%%%%%%%%%%%%%%% APPENDICES %%%%%%%%%%%%%%%%%%%%%

\appendix

\section{Additional Plots}

\begin{figure} 
\includegraphics[clip=,width=1\linewidth]{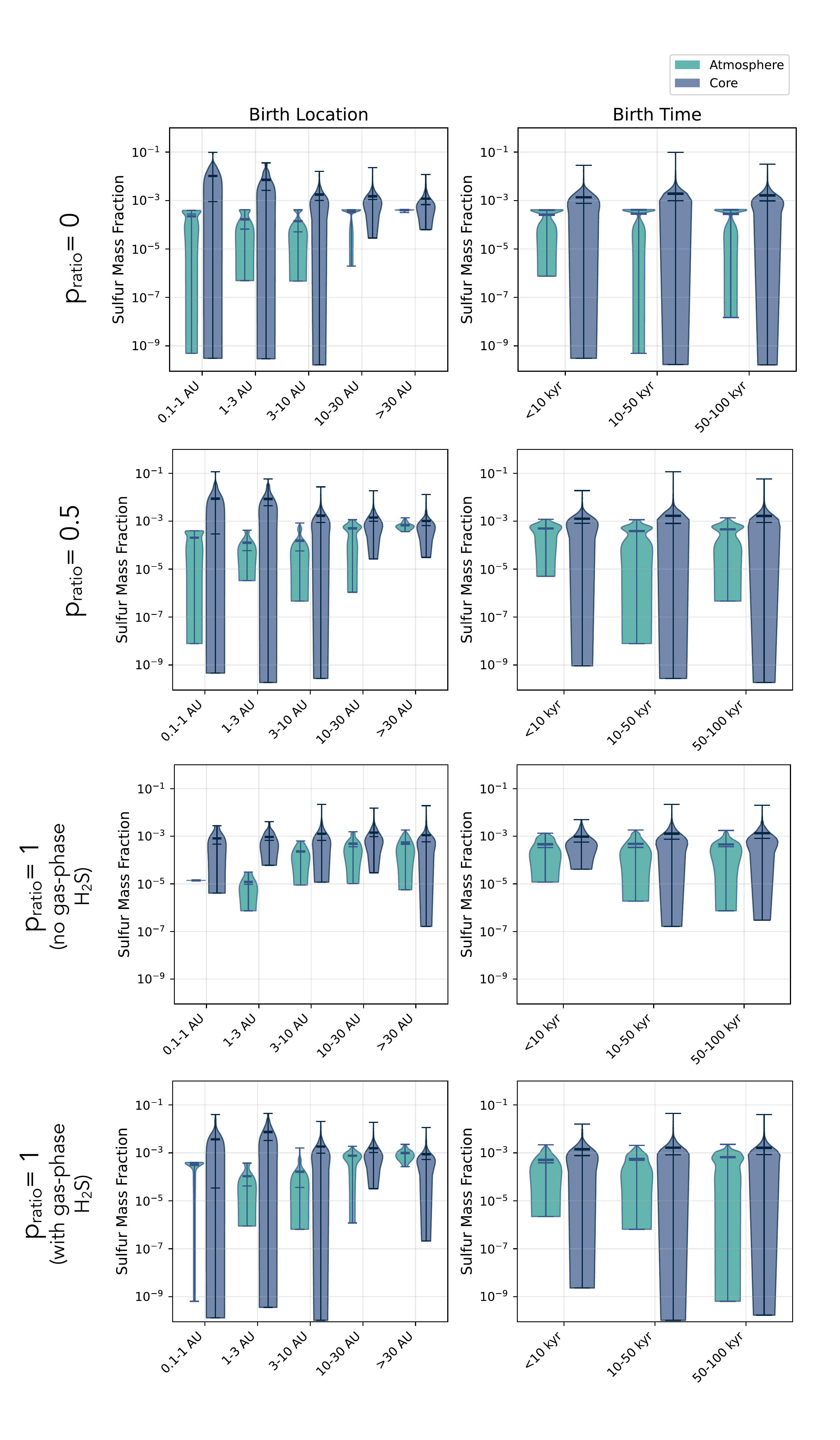}
\caption{Violin plots showing the non-zero sulfur mass fractions in the cores and envelopes of terrestrial planets and super-Earths ( $\leq10M_{\oplus}$) across all three disk models. Each row corresponds to one of the three values considered for the ratio of planetesimals ablated in the planets' atmospheres. Two cases of $p_{\rm ratio}=1$ are featured one where sulfur is only considered in the refractory carried by planetesimals, and one where the chemical kinetics described in Section \ref{sec:chem} are included, therefore including both gas-phase and refractory sulfur reservoirs. Jupiter's atmospheric sulfur mass fraction is marked in its appropriate mass category for reference.}
\label{fig:violinter}
\end{figure}

\begin{figure} 
\includegraphics[clip=,width=1\linewidth]{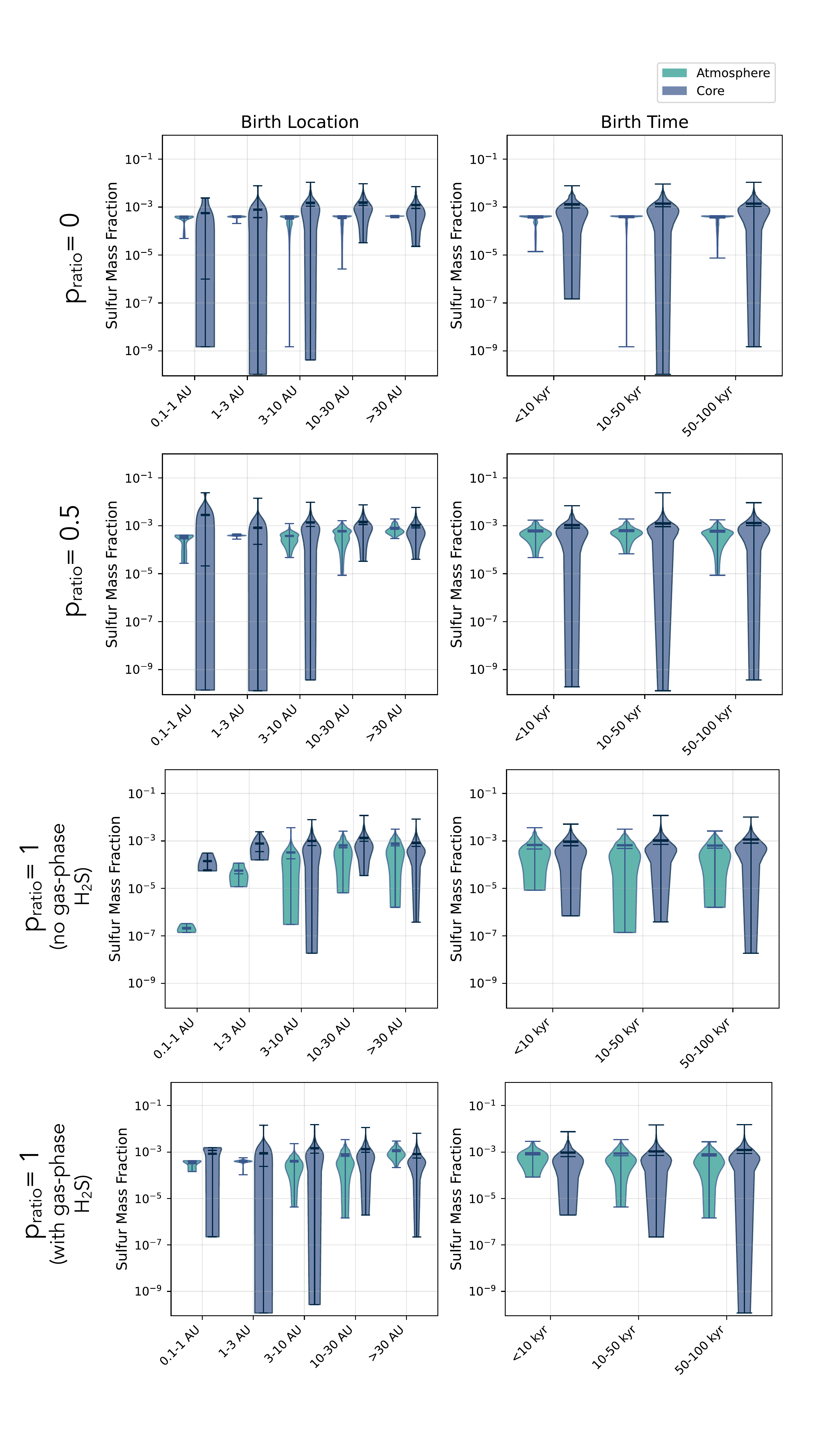}
\caption{Violin plots showing the non-zero sulfur mass fractions in the cores and envelopes of Neptunian planets ( $10 - 60M_{\oplus}$) across all three disk models. Each row corresponds to one of the three values considered for the ratio of planetesimals ablated in the planets' atmospheres. Two cases of $p_{\rm ratio}=1$ are featured one where sulfur is only considered in the refractory carried by planetesimals, and one where the chemical kinetics described in Section \ref{sec:chem} are included, therefore including both gas-phase and refractory sulfur reservoirs. Jupiter's atmospheric sulfur mass fraction is marked in its appropriate mass category for reference.}
\label{fig:violinnep}
\end{figure}

\begin{figure} 
\includegraphics[clip=,width=1\linewidth]{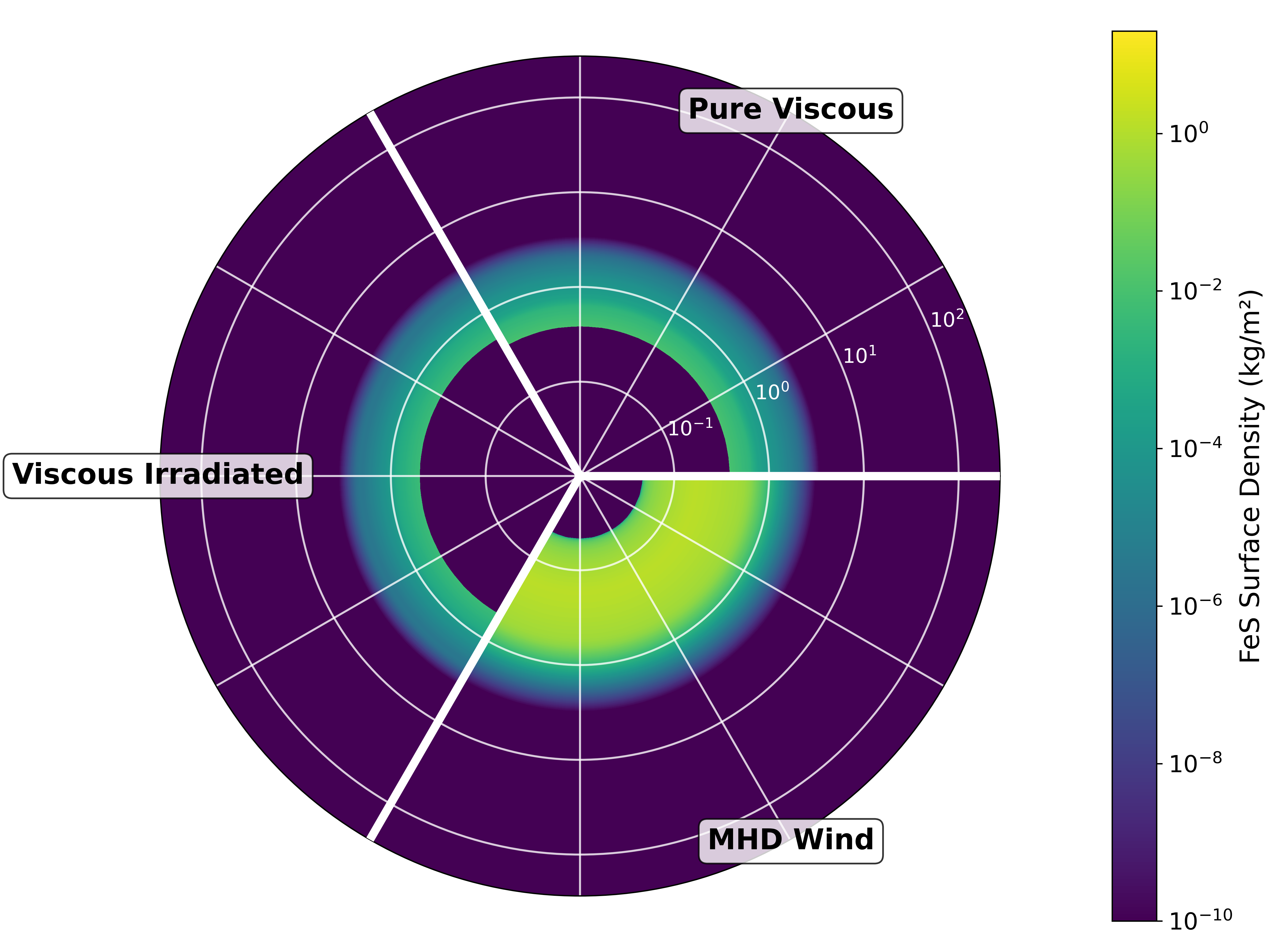}
\caption{Surface density distribution of FeS in each considered disk model after $3\,\text{Myr}$ of disk evolution without a planet present, using the fiducial disk parameters outlined in Table \ref{table:1}.}
\label{fig:h2s_sigma}
\end{figure}

%%%%%%%%%%%%%%%%%%%%%%%%%%%%%%%%%%%%%%%%%%%%%%%%%%

% Don't change these lines
\bsp	% typesetting comment
\label{lastpage}
\end{document}